\documentclass[fleqn,usenatbib]{mnras}

\usepackage{newtxtext,newtxmath}

\usepackage[T1]{fontenc}
\usepackage{ae,aecompl}
\usepackage{float}

\usepackage{graphicx}	
\usepackage{amsmath}	
\usepackage{amssymb}	
\usepackage{xspace}

\newcommand{\hmpc}{h^{-1}{\rm Mpc}}

\newcommand{\kms}{\;{\rm km}\,{\rm s}^{-1}}

\newcommand{\pygad}{{\sc Pygad}}
\newcommand{\simba}{\mbox{{\sc Simba}}\xspace}
\newcommand{\gizmo}{\mbox{{\sc Gizmo}}\xspace}
\newcommand{\mufasa}{\mbox{{\sc Mufasa}}\xspace}

\newcommand{\todo}[1]{\textcolor{black}{#1}}
\newcommand{\twodo}[1]{\textcolor{black}{#1}}

\title[X-rays in \simba]{X-ray Emission From Hot Gas in Galaxy Groups and Clusters in \simba}
\author[Robson \& Dav\'e]{Dylan Robson $^{1}$, Romeel Dav\'e $^{1,2,3}$
\\
\\$^1$ University of Edinburgh, Edinburgh, UK
\\$^{2}$ University of the Western Cape, Bellville, Cape Town 7535, South Africa
\\$^{3}$ South African Astronomical Observatories, Observatory, Cape Town 7925, South Africa
}
\date{Accepted XXX. Received YYY; in original form ZZZ}

\begin{document}

\label{firstpage}
\pagerange{\pageref{firstpage}--\pageref{lastpage}}

\maketitle

\begin{abstract}

We examine X-ray scaling relations for massive halos ($M_{500}>10^{12.3}M_\odot$) in the \simba galaxy formation simulation. The X-ray luminosity , $L_X$ vs. $M_{500}$ has power-law slopes $\approx\frac{5}{3}$ and $\approx\frac{8}{3}$ above and below $10^{13.5} M_{\odot}$, deviating from the self-similarity increasingly to low masses.  $T_X-M_{500}$ is self-similar above this mass, and slightly shallower below it. Comparing \simba to observed $T_X$ scalings, we find that $L_X$, $L_X$-weighted [Fe/H], and entropies at $0.1 R_{200}$ ($S_{0.1}$) and $R_{500}$ ($S_{500}$) all match reasonably well.  $S_{500}-T_X$ is consistent with self-similar expectations, but $S_{0.1}-T_X$ is shallower at lower $T_X$, suggesting the dominant form of heating moves from gravitational shocks in the outskirts to non-gravitational feedback in the cores of smaller groups.  \simba matches observations of $L_X$ versus central galaxy stellar mass $M_*$, predicting the additional trend that star-forming galaxies have higher $L_X(M_*)$.   Electron density profiles for $M_{500}>10^{14}M_\odot$ halos show a $\sim 0.1R_{200}$ core, but the core is larger at lower masses.  $T_X$ are reasonably matched to observations, but entropy profiles are are too flat versus observations for intermediate-mass halos, with $S_{\rm core}\approx200-400$~keV~cm$^2$. \simba's [Fe/H] profile matches observations in the core but over-enriches larger radii.  We demonstrate that \simba's bipolar jet AGN feedback is most responsible for increasingly evacuating lower-mass halos, but the profile comparisons suggest this may be too drastic in the inner regions.
 \end{abstract}

\section{Introduction}

As some of the largest gravitationally bound objects in the Universe, galaxy groups and clusters provide a unique laboratory for testing models of cosmology and galaxy evolution. The basic properties of groups and clusters are largely determined by the initial conditions and the hierarchical evolution of the dissipationless dark matter component, which is reasonably well understood. Since they probe the massive tail of the spectrum of mass perturbations, they are useful for constraining the growth of structure and hence the matter power spectrum.  But the formation of clusters also involves numerous baryonic processes that govern the appearance of the visible matter, in particular the galaxies and the hot virialised gas, which are less well understood.  Hence the co-evolution of galaxies, intergalactic gas, and black holes within groups and clusters allows us to investigate the physical processes that govern galaxy evolution in a unique environment.

During hierarchical collapse, adiabatic compression owing to gravity and subsequent shocks heat the intracluster gas to X-ray emitting temperatures ($T\ga 10^6 K$). In the simplest scenario, this gas then settles into hydrostatic equilibrium within the potential well of the cluster \citep{Kravtsov}.  For the most massive clusters, this configuration is stable, as the central cooling time is quite long~\citep{ReesOstriker_1977}.  For less massive systems, the gas in the central regions is expected to reach a sufficiently high density that it can radiatively cool on timescales short compared to the halo dynamical time.  The resulting cooling flow is then expected to feed star formation and accretion onto the super-massive black hole in the massive central galaxy. Simple estimates suggest that the cooling flows in clusters should be of order hundreds to thousands of solar masses per year, but most observed clusters show orders of magnitudes lower rates of stars formation, and very little gas cooling~\citep{Fabian:2002}.  This is known as the cluster cooling flow problem.

The most accepted scenario for solving the cooling flow problem invokes energetic feedback from active galactic nuclei (AGN) that counteracts cooling~\citep{Bower:2006,Croton:2006,Somerville,Somerville:2015}.  It is possible that magnetic conduction~\citep{NarayanMedvedev:2001,Fabian:2002} could play a role in preventing cooling flows, but cosmologically situated simulations suggest that it cannot fully balance the expected cooling rates~\citep{Jubelgas:2004}.  Supernova feedback could also provide some energy input, but since cluster galaxies are typically not star-forming there is little Type~II supernovae, and the Type~Ia supernovae are not sufficiently frequent to inject enough energy to prevent a cooling flow.
Observationally, clusters are seen to have bubbles of hot gas putatively blown by AGN jets, whose mechanical inflation work has approximately sufficient amplitude to counteract cooling~\citep{McNamara:2007}.  Hence jet-driven feedback from AGN are at present the leading candidate to inject large amounts of energy into the intracluster medium (ICM), and thereby quench star formation in massive galaxies and solve the cluster cooling flow problem.

Modeling the impact of AGN jet feedback within a cosmological structure formation context is challenging, owing to the large dynamic range involved and the relatively poor understanding of the physics of AGN feedback.  Nonetheless, recent improvements in simulation input physics along with the ever increasing computing power available have enabled a number of large-scale cosmological simulations that directly include black hole accretion and the resulting energetic output.  These include Illustris~\citep{Vogelsberger:2014,Genel:2014}, Magneticum~\citep{Hirschmann:2014}, Horizon-AGN~\citep{Dubois:2014,Volonteri:2016,Kaviraj:2017}, EAGLE~\citep{Schaye:2015}, MassiveBlack~\citep{Khandai:2015}, Blue Tides~\citep{Feng:2016}, Romulus~\citep{Tremmel:2017}, Illustris-TNG~\citep[TNG;][]{Springel:2018}, FABLE~\citep{Henden}, and \simba~\citep{Simba}. EAGLE, TNG, FABLE, and \simba\ were particularly successful at reproducing the observed massive red and dead galaxy population via AGN feedback, plausibly connecting AGN feedback with quenching of star formation and even morphological transformation~\citep{Genel:2014, Pillepich:2018, Dubois:2016}.  

While various simulations are now broadly successful at reproducing the galaxy population, they employ substantively different sub-grid models for black hole accretion and feedback. For instance, TNG and \simba\ directly employ high-velocity jet outflows from accreting black holes, FABLE provides input energy at the expected location of inflating bubbles, while EAGLE super-heats gas near the black hole to generate collimated AGN outflows.  In most cases, the AGN feedback model was specifically tuned to reproduce certain galaxy observations, such as the galaxy stellar mass function in which the high-mass exponential truncation is driven by AGN feedback.  To discriminate between these simulations and narrow down the plausible models of AGN feedback, we must thus rely on other observational diagnostics.  

One promising avenue is to look at the impact of AGN feedback on intragroup and intracluster gas.  \citet{McCarthy:2016} pointed out that the hot gas content of galaxy halos in the group and cluster regime can be a difficult observable for simulations to reproduce; this is still true even for otherwise successful models such as TNG~\citep{Barnes:2018}.  \simba\ and FABLE reproduce these data more closely, though not perfectly, with both FABLE and \simba\ slightly over producing hot gas in their most massive halos~\citep{Henden,Simba}.  Hence examining the hot X-ray emitting gas within groups and clusters appears to be a promising way to constrain the physical processes driving AGN feedback and galaxy formation in dense environments.

In this paper we examine the X-ray emission from hot halo gas in the \simba simulation. Specifically, we investigate X-ray scaling relations, and X-ray property profiles to determine the effectiveness of \simba's implemented feedback in reproducing observed trends. \simba\ employs three AGN feedback modules concurrently representing different physical modes of feedback, so by turning these on and off we are able to use several runs of \simba\ to test the impact of specific AGN feedback aspects. These varying runs include \simba\ which implements the full feedback model, \simba-NoX which includes stellar and jet feedback but no X-ray feedback, and \simba-NoJet which further turns off AGN jet feedback. We also compare to the older \mufasa\ simulation which does not include black holes and uses a less physically motivated heating model for quenching.  We find that the inclusion of jet feedback is mainly responsible for pushing halos away from self-similar predictions, and has by far the most important effect on both global properties and profiles when compared to other AGN feedback aspects, confirming expectations that AGN jets are crucial for reproducing observed hot gas properties in massive halos.

This paper is organised as follows: In \S2 we discuss the simulation code, and more specifically the AGN feedback mechanisms implemented,  we also discuss the variants of AGN feedback employed here, and outline how the X-ray emission is computed. In \S3 we discuss the self-similar scaling relations of halos. In \S4 we investigate the cluster mass budget by looking at the mass fractions of halos within the \simba simulation. In \S5 we discuss the mass scaling relations of \simba halos and also the comparisons between scaling relations within \simba\ , \simba NoX, \simba No-Jet, and \mufasa. In \S6 we investigate the X-ray radial profiles in \simba. 
Finally in \S7 we summarize our results.

\section{Simulations and Analysis}

\subsection{The \simba simulation}

\simba \citep{Simba} is a cosmological hydrodynamic simulation run using the \gizmo code. The \simba simulation models a $(100h^{-1}$Mpc$)^{3}$ random cosmological volume  with $1024^3$ dark matter particles and $1024^3$ gas elements evolved down to $z=0$.  In accord with \citet{Planck_Cosmology} it adopts a $\Lambda CDM$ cosmology with $\Omega_{\Lambda}  = 0.7,\Omega_m  = 0.3,\Omega_b  = 0.048, h = 0.68, \sigma_8 = 0.82 $, and $n_s = 0.97$. More information on the details of \simba can be found in \citet{Simba}; here we briefly recap key modeling elements.

\simba\ implements star formation using an $H_2$-based model, where the molecular fraction is computed following the prescription of \citet{KrumholzGnedin2011}.
Chemical enrichment is followed for 9 metals ejected owing to Type II and Type Ia supernovae and asymptotic giant branch (AGB) stars.
{\sc Grackle-3.1} is used for radiative cooling and photoionisation heating assuming a \citet{HaartMadau} ionising background.  We note that \simba, like its predecessor \mufasa, includes energy input from AGB stellar winds in the form of heating of the surrounding gas~\citep{Mufasa}, which \citet{Conroy2014} argued could be an important preventive feedback mechanism in early type galaxies.  

Galactic winds from star formation are modelled using decoupled two-phase winds, with a mass loading factor that scales with galaxy stellar mass $M_*$ as predicted by the tracking of individual particles in the FIRE simulations \citep{2017MNRAS.470.4698A}.  The wind velocity, implemented as a kinetic kick to the gas elements, scales approximately with the host galaxy circular velocity.  Decoupled winds means that a gas element launched in an outflow does not interact hydrodynamically for some time, until it reaches a density that is 1\% of the star formation density threshold, or has been a wind for 2\% of a Hubble time (at launch).  Galaxy properties such as $M_*$ are computed on-the-fly via an approximate friends-of-friends (FOF) finder applied to stars and dense gas \citep{Dave_2016}. This FOF technique also allows for black holes to be seeded on the fly by calculating galaxy properties as the simulation is run. Halos are also identified using a 3D FOF finder with a linking length of 0.2 times the mean inter-particle spacing.  

Black holes are seeded at $10^4M_\odot$ when galaxies exceed $10^{9.5}M_\odot$ in stellar mass, and grown in two modes: A mode applicable for cold gas where angular momentum loss is the primary bottleneck to accretion \citep[torque-limited accretion;][]{TLA} for gas with $T<10^5$K, and \citet{Bondi} accretion at higher temperatures.  \simba\ also includes AGN feedback; owing to its central nature in this work, we describe this implementation in the next section. \todo{Simulations from the FIRE project found that stellar feedback strongly suppresses black hole growth in low mass galaxies, motivating the galaxy stellar mass threshold used in \simba \citep{Angles:2017c}.}

For this analysis focusing on hot halo gas, we mostly restrict our study to halos with a halo mass $M_{500}>10^{12.3} M_{\odot}$. This results in a total of 1379 halos above this mass limit in the $100h^{-1}$Mpc$^3$ volume at $z=0$.  We will further consider higher mass limits, closer to what is observable with current X-ray telescopes, and employing an $M_{500}$ limit.  In these cases, there are 229 halos above $M_{500}>10^{13} M_{\odot}$ and 9 halos $>10^{14} M_{\odot}$, with a maximum halo mass of $10^{14.8} M_{\odot}$. 

\subsection{AGN feedback in \simba}

Due to its highly energetic nature, AGN feedback can have a significant impact on the host galaxy and its evolution. Active galactic nuclei (AGN) feedback can be split into two main modes, as described in \citet{Heckman}: Radiative mode at high Eddington ratios ($f_{\rm Edd} \equiv \dot M_{\rm BH} / \dot M_{\rm Edd}$); and jet mode at lower $f_{\rm Edd}$.  Radiative mode AGN have their energetic output dominated by electromagnetic radiation emitted by the accretion disk around the central super massive black hole (SMBH).  AGN jets have energetic output that is dominated by bulk kinetic energy in the form of collimated jets, powered by gas accretion and/or from the spin of the SMBH.  There is also the impact of photon pressure from high-energy (X-ray) radiation generated by the accretion disk.  \simba\ thus incorporates three types of black hole feedback:  Radiative winds, jets, and X-ray photon pressure.  

Radiative winds are thought to arise when the black hole has a cold accretion disk that reaches into the innermost stable circular orbit. The energetic output from the disk results in photon pressure that lifts material off the disk, up to speeds of $\ga 1000\kms$.  By entraining surrounding material, the total momentum input can be an order of magnitude higher than $L/c$, where $L$ is the AGN luminosity.  Such winds are observed as ionised~\citep[e.g.][]{Perna:2017} or molecular~\citep[e.g.][]{Sturm_2011} outflows, typically detected as broad emission line wings.

At low accretion rates, the black hole accretion changes character, with the black hole now being surrounded by a hotter torus whose accretion is advection dominated.  This is thought to occur at accretion rates below around $1-2\%$ of the Eddington rate~\citep{BestHeckman:2012}.  The angular momentum and magnetic field of the accretion flow drive a highly collimated relativistic jet of high-energy particles out from the polar directions.

Black holes also produce significant X-ray emission off their accretion disks, in either mode.  This provides photon pressure on material surrounding the black hole.  When the immediate environment is cold gas-rich, it is likely that any of the X-ray photon pressure that gets absorbed is quickly radiated away by the dense gas.  However, at low accretion rates when the surrounding gas is hotter, the photons can provide a net outwards momentum, which is more spherical in nature than the jet feedback.

\simba includes each of these modes of AGN feedback, in a manner that attempts to mimic observations as closely as feasible.  The interplay of all of these black hole feedback modes with the surrounding gas provide key feedback mechanism that significantly affects the growth and evolution of their host galaxies as well as the gaseous halos in which they sit.We now describe these models.

To model radiative and jet feedback in \simba, we employ purely kinetic and bipolar, \todo{continuous} outflows, acting parallel to the axis of angular momentum within the inner disk, defined by the 256 nearest neighbours to the black hole. For radiative mode, we set the outflow velocity based on observations of the ionised gas linewidths of X-ray detected AGN (\citet{Perna:2017}), parametrised in terms of black hole mass $M_{\rm BH}$ as:
\begin{equation}
    v_{\textrm{w,EL}} = 500 + 500(\log M_{\textrm{BH}} - 6)/3 \kms  
\end{equation}

As $f_{\rm Edd}$ drops to $<0.2$ the jet feedback begins to add an additional velocity component, whose strength depends on the Eddington ratio:
\begin{equation}
    v_{\textrm{w,jet}} = v_{\textrm{w,EL}} + 7000 \log(0.2/f_{\rm Edd}) \kms ,
\end{equation}
with the velocity increase capped to $7000\kms$ at $f_{\rm Edd}$\la 0.02.  The logarithmic dependence means that the velocity ramps up more quickly towards low $f_{\rm Edd}$.  Hence full jet mode is achieved only below a couple percent of Eddington, at speeds of $\sim 8000 \kms$. \todo{These gas wind elements experience a short hydrodynamic and radiative cooling decoupling time, and as such are not subject to significant radiative losses in the dense gas surrounding the black hole.} 

X-ray feedback is simulated using a spherical input of kinetic energy into the gas surrounding black hole if it is star forming-gas, or else thermal input if non-star forming.  The kinetic input is required because in \simba\ the star-forming gas is forced to lie on a density--temperature relation as specified to resolve the Jeans mass~\citep[see][]{Dave_2016}, so thermal input would have limited effect. X-ray feedback is only active alongside full velocity jets ($f_{\rm Edd}<0.02$), since in this case the immediate surroundings are expected to be free of cold gas that would radiatively cool away the energy input.  We further require the galaxy have a cold gas fraction $f_{\rm gas} < 0.2$, to model the assumption that gas rich galaxies are able to absorb and radiate away X-ray energy.

In order to isolate the physics responsible for setting the various scaling relations, we run alternative versions turning off various AGN feedback forms.  Owing to computational limits, we run these in a $50h^{-1}$Mpc$^3$ box with $512^3$ dark matter particles, and $512^3$ gas elements, thus having the same resolution as our full $100\hmpc$ run but with $8\times$ less volume.  Unfortunately, this reduces the number of high mass halos in these volumes, nonetheless we can still glean some interesting trends.  Specifically, we run a ``no-jet" model (\simba-NoJet) where we turn off both the jet mode and X-ray feedback, and a ``no-X" model (\simba-NoX) where we only turn off the X-ray feedback but leave jets on.  In all cases the radiative AGN feedback remains on, but radiative feedback is found to have very minimal impact on galaxy or halo gas properties, so for clarity we do not include this because the results is very similar to the ``no-jet" case.  Finally, as a point of comparison, we also include results from the \mufasa\ simulation~\citep{Dave_2016}, which used a halo heating model as a proxy for AGN feedback, also in the same box size.  The initial conditions for all these $50\hmpc$ runs are identical.

\subsection{Computing X-ray emission} \label{Xray}
\begin{figure}
    \centering
    \includegraphics[width = 0.45\textwidth]{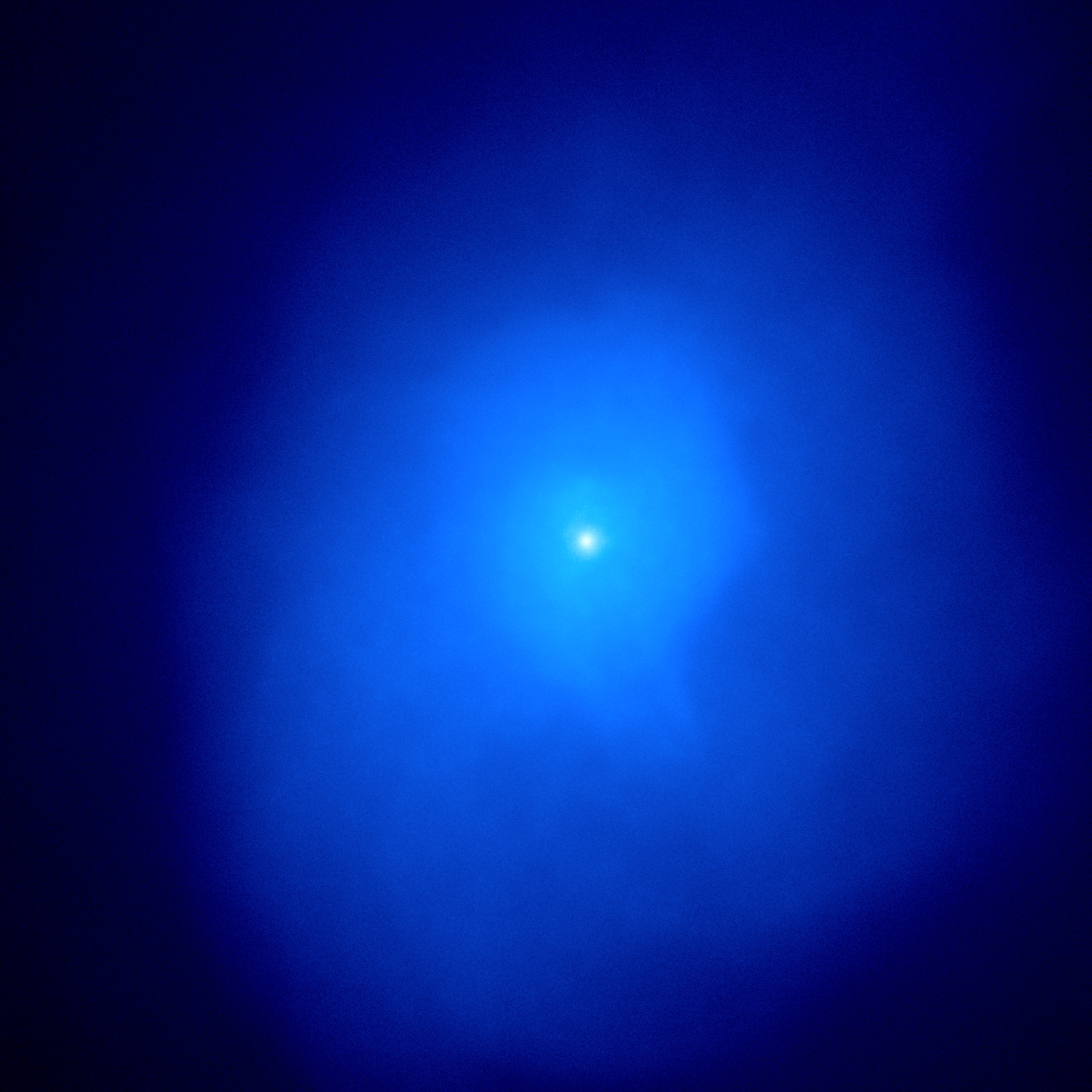}
    \caption{X-ray photons produced by the most massive halo in \simba at $z = 0$, with an $M_{500}$ of $6.6 \times 10^{14} M_{\odot}$. \todo{This image was created using pyxsim and SOXS and and spans a physical distance of $1.3$~Mpc on a side.}}
    \label{fig:my_halo}
\end{figure}

X-rays are computed using the python module \pygad\ \citep{pygad}, a multipurpose tool that allows for the general analysis of Gadget-based simulations. \pygad\ allows for the creation of a sub-snapshot based on various criteria such as desired particles, specific regions such as individual FOF halos, or particles within specific property (e.g. temperature or metallicity) range.

\pygad\ includes a an X-ray luminosity analysis module that utilises XSPEC \citep{XSPEC}, an X-ray spectral fitting package, to compute the X-ray spectrum, described more fully in \citet{Eisenreich_2017}.  By using the pre-prepared emission tables from XSPEC \pygad\ is able to calculate the X-ray luminosity of selected gas particles based on the particle temperatures and metallicities.  In our analysis of \simba's X-ray properties we use the provided 0.5--2 keV X-ray table which is similar to the range quoted for most of the observations we will compare to.  \pygad\ is publicly available at {\tt https://bitbucket.org/broett/pygad}.

In Figure \ref{fig:my_halo} we show events created from X-ray photons produced using XSPEC within the most massive halo in \simba. The halo has $M_{500}$ of $10^{14.8} M_{\odot}$ and an X-ray luminosity of roughly $10^{45}$ erg s$^{-1}$. We see a centrally peaked luminosity with a relatively smooth, if somewhat asymmetric, decline towards the edges. This is usual for the most massive halos, typically being dominated by a single central galaxy, while lower mass halos tend to contain satellite galaxies with luminosities approaching those of their respective central galaxy. 

\section{Self-similar scaling relations}

Hot gas appears in galaxy halos primarily owing to shock heating of gas as massive objects collapse during gravitational structure formation~\citep{ReesOstriker_1977}.  Such shock heating happens only in halos above $M_{\rm halo}\ga 10^{11.5-12}M_\odot$, as below this mass the accretion shock is unstable due to radiative cooling \citep{DekelBirnboim_2003, Keres_2005, Gabor:2012,Nelson:2013}.  Hence X-ray emission from hot halo gas generally only appears in fairly massive systems.

Assuming only gravity and free-free emission from the hot gas, it is possible to derive expected scaling relations between the halo mass and the X-ray emission. These are known as the self-similar scaling relations~\citep{Kaiser}, because in this case there is no intrinsic scale and hence halos of all masses are self-similar.  Real groups and clusters are known to deviate from self-similarity, through gas cooling out of the hot phase via radiative processes~\citep{voit2005expectations}, as well as feedback providing additional energy to gas.  The self-similar relations thus provide a baseline from which to measure deviations owing to non-gravitational processes.

For a given virialized density contrast $\Delta\approx 200$, the mass of a halo is given by
\begin{equation}
M_{\Delta z} = \frac{4 \pi}{3} \Delta \rho_{\rm crit,0} E_z^2 R^3 
\end{equation}
where $R$ is the halo radius, $\rho_{\rm crit,0}$ is the $z=0$ critical density, and $E_z = H_z/H_o = [(\Omega_m ( 1 + z)^3 + ( 1 - \Omega _m - \Omega _{\Delta})(1+z)^2 + \Omega_{\Delta})]^{1/2}$
describes how the Hubble parameter evolves with $z$. In hydrostatic equilibrium, the thermal energy of the hot gas must balance the gravitational potential energy, hence
\begin{equation}
T_{X} \propto \frac{GM}{R} \propto R^2
\end{equation}
where $T_X$ is the hot gas temperature. Therefore our first self-similar scaling relation relates halo mass and X-ray temperature:
\begin{equation}
M \propto T_{X}^{3/2}.
\end{equation}
To obtain the expected X-ray luminosity, we must appeal to free-free emission (thermal bremsstrahlung) as the dominant emission mechanism. \todo{For systems in which the ICM has been heated to $\ga 10^6$ t}he free-free volumetric emissivity is given by
\begin{equation}
    \epsilon \approx 3\times 10^{-27} T_{X}^{1/2} \rho_{\rm gas}^{2}\;\; {\rm erg\; cm}^{-3} {\rm s}^{-1}.
\end{equation}
where $\rho_{\rm gas}$ is the gas density  \citet{Giodini:2013}. If we assume that the density profile of all halos are self-similar, then the X-ray luminosity scales as 
\begin{equation}
    L_X \propto \epsilon R^3  \propto T_X^{1/2} \rho_{\rm gas}^2 R^3 \propto T_X^{1/2} f_{\rm gas}^2 R^3 \propto f_{\rm gas}^2 T_X^2,
\end{equation}
and thus for a constant gas fraction, $f_{\rm gas}$, 
\begin{equation}
    L_X \propto T_{X}^2 \propto M^{4/3}.
\end{equation}
Finally, in X-ray studies it is common to define the entropy $S_X\equiv T_X/n_e^{2/3}$.  For self-similarity, $n_e$ is independent of mass, so $S_X\propto T_X\propto M^{2/3}$.

This gives the expected scaling for X-ray luminosity versus temperature, mass, and entropy under the assumption of self-similarity \citep{Vikhlinin,Kravtsov:2012}. These scaling relations hold for halos of gas experiencing solely gravitational heating processes.  As a result, departures from self similarity can be evidence of non-gravitational processes taking place, such as radiative cooling and AGN feedback.  

\section{Baryonic Mass Budget}

Energy input from AGN within \simba quenches massive galaxies in large halos.  The main feedback mode responsible for quenching is AGN jets~\citep{Simba}.  Jets are able to directly evacuate some gas from halos owing to their large velocities, but also deposit heat into halo gas in such a way that it becomes unbound from the halo.   As a result, AGN feedback has a strong impact on the hot gas within halos.  The most basic quantification of this is the amount of baryons remaining in the halo.  Hot gas fractions are seen to deviate from the cosmic mean baryon fraction particularly at group scales, as observed via X-ray emission from intra-group and intra-cluster gas. Also, stellar baryon fractions are increasingly reduced towards massive halos relative to $\sim 10^{12}M_\odot$ halos where they they peak~\citep{Moster:2013,Behroozi:2013}. These observations provide important constraints on the AGN feedback model.

In this section we examine the baryonic mass budget in stars and hot gas within massive \simba\ halos.  Unless otherwise specified, we will focus on halos with $M_{500}>10^{13}M_\odot$, which is the regime where X-ray data on diffuse halo gas is generally available.  We will focus on computing quantities to $R_{500}$, which is the radius out to 500 times the critical density computed via a spherical overdensity algorithm, since that is commonly what is quoted from X-ray observations. Broadly, the results are similar to those within $R_{200}$, but some quantities are slightly systematically biased by focusing on the inner $\sim 2/3$ of the virial radius.

\subsection{Stellar mass fraction}

\begin{figure*}
    \centering
    \includegraphics[width = 0.49\textwidth]{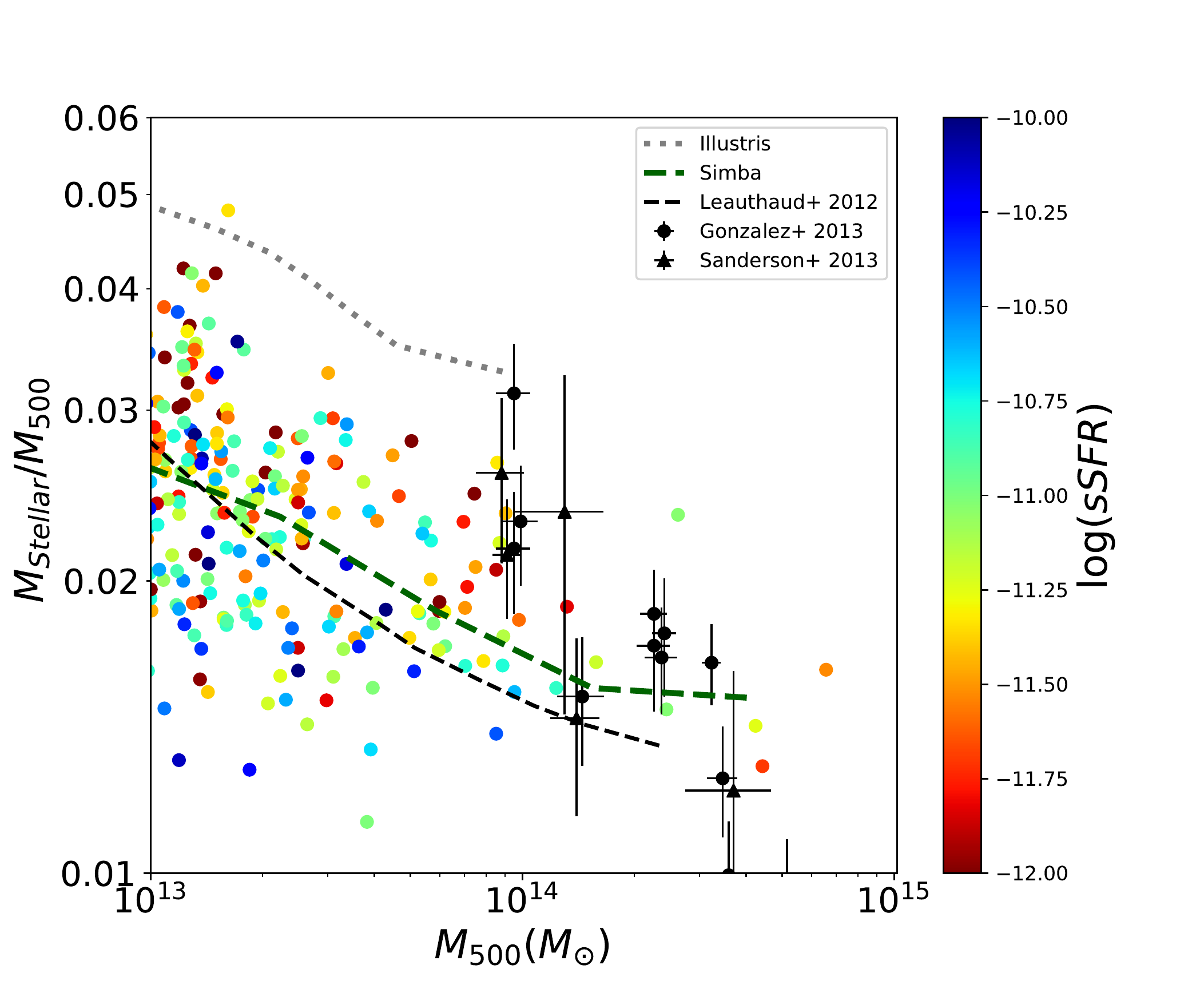}
    \includegraphics[width = 0.49\textwidth]{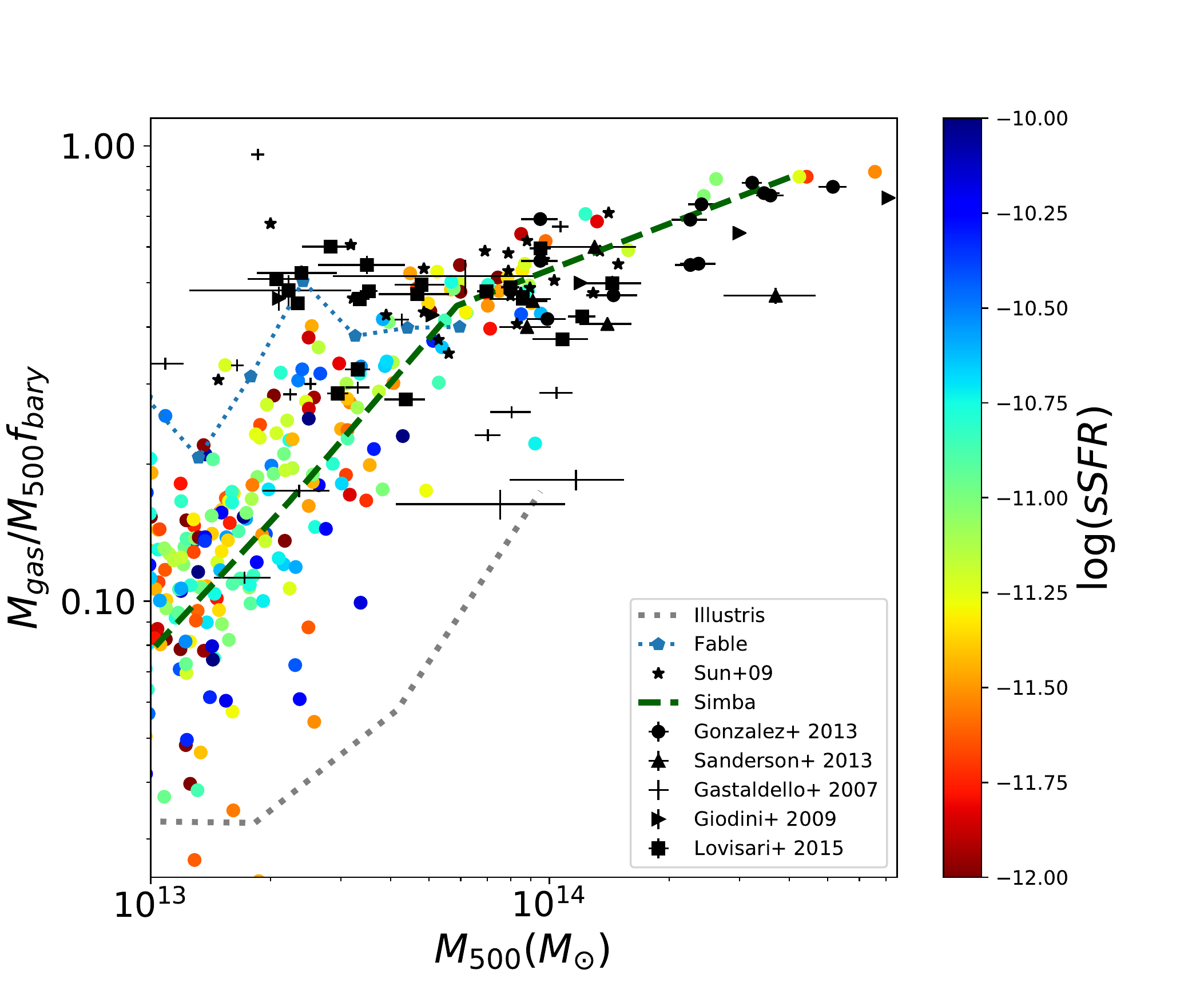}
    \caption{Stellar (left) and hot gas (right) mass fraction of clusters at $z=0$ in the $100 h^{-1}$Mpc$ ^3$ \simba box, with colour scaled by the total specific star formation rate out to $R_{500}$. Observations of the stellar mass fraction are shown from X-ray data of \citet{Gonzalez} and \citet{Sanderson};  following \citet{Chiu} and \citet{Henden}, these have been reduced by 24\% to change from Salpeter to our assumed Chabrier IMF.  We show a fit to weak lensing-based measurements from \citet{Leauthaud} as the dashed black line.  The dotted line is the mean relation taken from Illustris \citep{Genel:2014} at $z=0$. \simba shows good agreement with the X-ray observations at the highest $M_{500}$, and agrees more closely with \citet{Leauthaud} towards the lower mass halos. In the right panel, observations of the hot gas mass fraction come from \citet{sun9}, \citet{Gonzalez}, \citet{Sanderson}, \citet{Gastaldello}, \citet{Giodini}, and \citet{Lovisari}. The blue dotted line is the mean relation from FABLE~\citep{Henden}, while the black dotted line is the mean relation from Illustris (\citet{Genel:2014}). \simba largely agrees with observations, sitting on the high end for $M_{500} > 10^{14} M_{\odot}$. \simba produces reasonable agreement with observations of both stellar and gas content in galaxy groups, suggesting that it offers a plausible platform to investigate X-ray properties of intragroup gas.}
    \label{fig:StellarMass}
\end{figure*}

Figure \ref{fig:StellarMass}, left panel, shows the stellar-to-$M_{500}$ ratio within $R_{500}$ of all halos above $10^{13} M_{\odot}$, as a function of $M_{500}$.  Here we include all stars, not just those in the central galaxy, though typically the central galaxy dominates the mass.  The points are color-coded by the halo specific star formation rate (sSFR), computed as the total star formation rate over the stellar mass in all galaxies out to $R_{500}$. The running median is shown as the dashed green line.  At the massive end, X-ray determined observations are shown from \citet{Gonzalez} and \citet{Sanderson}, including the contribution of intracluster light (ICL) which can be very significant at the low mass end.  Also, the black dashed line shows observations from a weak lensing plus halo occupancy distribution analysis from \citet{Leauthaud}, which probes to lower masses than can be done via X-ray emission.  Finally, for comparison, we show the mean relation from the Illustris simulation as the grey dotted line.

The \simba\ halos broadly match the data from \citet{Gonzalez} and \citet{Sanderson} for the massive halos, falling slightly below these observations at $M_{500} \la 10^{14}M_{\odot}$. In this regime, the contributions from the ICL are fairly strong, but are relatively more uncertain.   \citet{Gonzalez} points out that previous determinations of this ratio have found somewhat shallower trends of stellar fraction versus $M_{500}$, which they attribute to not including the ICL contribution.

A shallower slope is seen for the \citet{Leauthaud} data, and here \simba\ is in better agreement at $M_{500} \sim 10^{13}M_{\odot}$, as the trend with $M_{500}$ in these observations is significantly shallower than seen for the X-ray data.  The disagreement between the trends was pointed out in \citet{Gonzalez}, and it is beyond the scope of the present work to assess this; it could be that the HOD analysis fails to accurately account for the ICL, or it could be that the ICL is overestimated owing to foreground/background contamination.  A more careful comparison mimicking each dataset is warranted, but we leave that for future work.

Finally, examining the colours of the plotted \simba\ halos, we do not detect any systematic trend in the stellar fraction with the sSFR within the halo.  The sSFR will correlate with other properties we examine later, so we show it here for reference. This suggests that the high stellar fraction halos do not have systematically higher current sSFR of their member galaxies; instead, the high stellar mass must be related to its growth history.  Similarly, we found no obvious trend versus current black hole mass or accretion rate (not shown).  We leave for future work an investigation of what determines the spread in stellar fractions at a given halo mass, via tracking halos back over time.

We conclude that the predictions of \simba\ for stellar baryon fractions seem to be in accord with available observations for more massive ($\ga 10^{14}M_\odot$) halos, though may under-predict the stellar fraction in lower mass halos when observations account for the ICL, with the caveat that such observations come with larger systematic uncertainties. It is notable that the amplitude agreement with observations is not a trivial outcome of galaxy formation simulations, as evidenced by the fact that Illustris shows ratios that are well higher. However we note \simba was tuned to the galaxy stellar mass function, while Illustris was not.  Overall, within the range probed by X-ray observations ($M_{500}\ga 10^{13.5}M_\odot)$, \simba generally produces reasonably good agreement with stellar baryon fractions given the uncertainties.

\subsection{Hot gas mass fraction}

In galaxy clusters, the majority of baryons are in the form of hot gas; in groups, this is less clear.  As mentioned earlier, the hot gas fraction within groups and clusters has been a challenging observations for simulations to reproduce~\citep{McCarthy:2016}.  In \citet{Simba} we presented a preliminary prediction of this, but here we update this via a more careful comparison to a fuller suite of observations.

Figure \ref{fig:StellarMass}, right panel, shows the hot gas mass fraction within $R_{500}$ of all halos above $10^{13} M_{\odot}$.  We define hot gas as having $T\geq 10^{5.5}$~K, but these results are generally insensitive to any choice from $10^5-10^6$~K.  The dashed green line shows the running median from \simba.  The points are colour-coded by overall halo specific star formation rate, as in the left panel.  Observations are shown from \citet{Gonzalez}, \citet{Sanderson}, \citet{Gastaldello}, \citet{Giodini}, and \citet{Lovisari}. For model comparisons, the running median from Illustris is shown as the dotted grey line, and the running median from FABLE is shown as the dotted blue line.

This plot is similar to Figure~8 in \citet{Simba}, but there the hot gas fraction was computed over the entire halo (although plotted vs. $M_{500}$) whereas here we have computed the hot gas fraction specifically within $R_{500}$.  Hence it is a more accurate comparison to the observations.  This tends to slightly increase the hot gas fraction, as hot gas tends to be more centrally concentrated in galaxy groups.  Relative to their plot, we also show the observations as individual data points, to illustrate the scatter in observations more clearly.

At the massive end, \simba\ has 80-90\% of its halo baryon fraction in the form of hot gas.  These numbers are typical of observed galaxy clusters.  There is a weak dropoff towards smaller halos down to $\sim 10^{14}M_\odot$, and then the dropoff steepens towards lower masses, with a substantial increase in the scatter.  This suggests that stochastic AGN feedback processes have a larger impact in poor group-scale halos than in the most massive systems.

\simba\ produces generally good agreement with the observations, both qualitatively and quantitatively.  However, the \simba\ halos tend to lie at the high end of the observations for $M_{500}\ga 10^{14}M_\odot$, indicating a potential discrepancy.  We note that our temperature limit of $10^{5.5}$K is well below what can typically be traced by X-ray observations, so it's possible that the \simba\ values are biased high because of this.  Nonetheless, this level of agreement is a non-trivial success, as for instance shown by the stronger disagreement with Illustris.  More recent simulations tend to show agreement at a similar level to that seen in \simba~\citep[e.g. Illustris-TNG and FABLE;][]{Barnes:2018,Henden}.  \simba\ agrees well with FABLE for their $M_{500}\ga 10^{13.5}M_\odot$, but FABLE tends to show a less steep relation to lower masses; current observations have a large scatter in this regime that encompasses both models' predictions.

In summary, \simba\ produces stellar and hot gas baryon fractions that are in reasonable agreement with available observations.  It is worth noting that no specific tuning of the feedback model in \simba\ was done in order to match hot gas properties in galaxy groups, as the tuning focused on the overall stellar mass function and black hole properties.  Hence this level of agreement is encouraging, and indicates that \simba\ provides a plausible platform to investigate the X-ray properties of intra-group gas, as we do next.

\section{X-Ray Scaling Relations}

The most basic X-ray observation for a group or cluster of galaxies is its total X-ray luminosity. The comparative ease with which X-ray luminosities can be measured enables X-ray surveys to estimate the mass of clusters, making X-ray luminosity--mass relation a crucial tool for cosmology. Studies have found that the observed $L_X - M_{500}$ slope is $\sim 1.4 - 1.9$, suggesting non-gravitational processes push the relation away from the slope of $4/3$ as expected from self-similarity.  Measurements of the X-ray temperature $T_X$ versus mass likewise show deviations from self-similarity, though typically only mildly.  Hence such X-ray scaling relations allow us to probe the gas and galaxy physics at play in massive halos.  

Hot gas metallicity and entropy further highlight the impact of cluster galaxies, where the metals are generated, and their feedback processes that distribute metals and alter the gas entropy.  The hierarchical formation history and resulting gas dynamics can introduce a significant scatter into these relations, with much of the scatter being introduced in the core of the cluster where effects such as merging and cooling are more pronounced.  Thus X-ray scaling relations embody a complicated set of processes that connect structure formation and galaxy growth, and are a key benchmark for galaxy formation models. 

In this section we examine X-ray property scaling relations versus $M_{500}$ and $T_X$ in \simba, to quantify the impact of such non-gravitational processes within groups and clusters.  The two comparisons are closely related, but the first is more theoretically oriented and connects better with the amounts and properties of the hot gas, while the latter is more observationally oriented and enables a wider suite of data comparisons to more carefully test the simulation.

\subsection{Scaling Relations Versus Mass}

\begin{figure*}
    \centering
    \includegraphics[width = \textwidth]{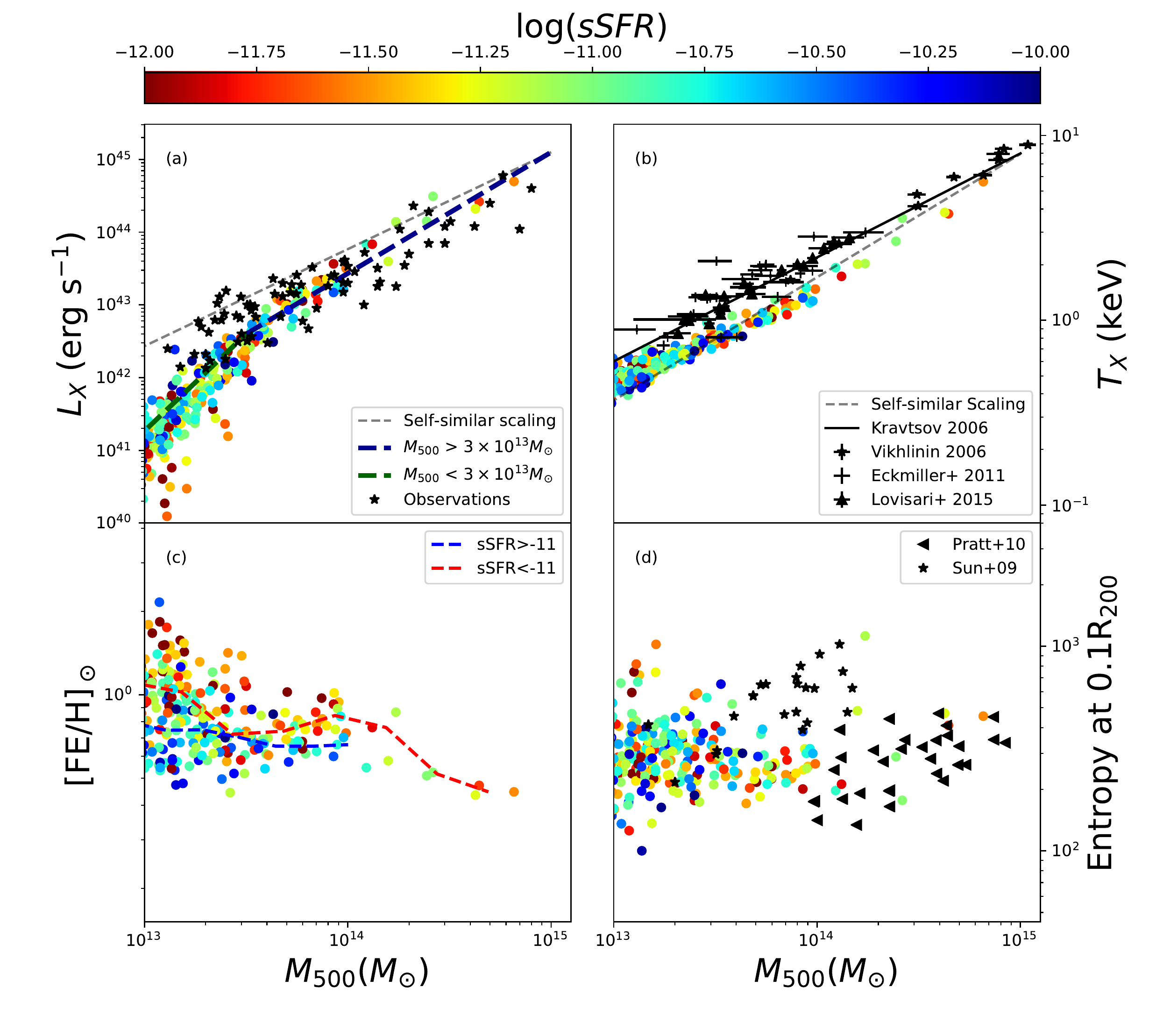}
    \caption{X-ray scaling relations versus halo mass $M_{500}$ of halos in \simba, with points colour-coded by halo sSFR.  (a) The top left subplot shows the $L_X-M_{500}$ relation, compared with observations from \citet{Vikhlinin},\citet{Pratt}, \citet{Eckmiller}, \citet{Lovisari} and  \citet{sun9} indicated by black stars.  Self-similar scaling is shown as thin dashed line, and power law fits of \simba data below and above $M_{500}=3\times 10^{13} M_{\odot}$ are also plotted with slopes of $8/3$ and $5/3$, respectively.
    (b) The top right subplot shows the X-ray luminosity weighted temperature $T_X$ vs. $M_{500}$, compared with the observed fit taken from \citet[black line]{Kravtsov}, and observations from \citet{Vikhlinin},\citet{Eckmiller}, and \citet{Lovisari} plotted over. The self-similar relation is also plotted as the dashed black line, showing \simba halos largely follows this relation, deviating slightly at the lowest masses. (c) The bottom left subplot shows the $L_X$-weighted iron abundance [Fe/H] vs. $M_{500}$.  We show running medians plotted in blue and red for sSFR $> -11$ and sSFR$<-11$ respectively, showing that low-sSFR halos tend to have higher hot gas metallicity at a given $M_{500}$. (d) The bottom right subplot shows the relation of $M_{500}$ versus $S_{0.1}$, the X-ray weighted entropy at $0.1R_{200}$, with observational data from \citet{Pratt:2010}. \simba halo lie in the range of observations, and show little to no trend with $M_{500}$.}
    \label{fig:Mass_Scaling}
\end{figure*}

Figure \ref{fig:Mass_Scaling} shows the scaling relations of a X-ray luminosity $L_X$ (panel a), X-ray luminosity weighted temperature $T_X$ (b), X-ray luminosity weighted iron metallicity (c), and the X-ray weighted entropy at 10\% of the virial radius, $S_{0.1}$ (d), versus $M_{500}$ for $>10^{13}M_\odot$ halos at $z=0$ within \simba. Entropy $S_{0.1}$ is calculated as $T_X/n_e^{2/3}$, using particles located between $0.05 R_{200}$ and $0.15 R_{200}$.  The points are colour coded by halo specific star formation rate. For $L_X-M_{500}$, observational data from \citet{Vikhlinin}, \citet{Pratt}, and \citet{sun9} are plotted as stars. For $T_X-M_{500}$, observational data from \citet{Kravtsov}  and \citet{Vikhlinin} is plotted. Observational data from \citet{Pratt:2010} is plotted for entropy at $0.1 R_{200}$.

Comparing the $L_X-M_{500}$ relation to observational data in panel (a) shows that \simba halos broadly follow observed trends.  The predicted $L_X-M_{500}$ relation follows a slope of $\approx 5/3$ down to $M_{500}\approx 10^{13.5} M_\odot$ where it experiences a break and follows a slope of $\approx 8/3$. This break is suggestive of the effect of AGN feedback on halo gas in which gravitational heating becoming dominant over feedback heating towards more massive halos.  This is also seen in the hot gas fractions which have a steep dependence on $M_{500}$ below this mass (see Figure~\ref{fig:Fracs}).

There is still a mild deviation from self similarity even at the most massive halos in \simba, which is also indicated by the observations.  The most massive clusters at $M_{500}>10^{15}M_\odot$ are expected to follow self similarity, but our $100\hmpc$ volume is  too small to contain such systems. \todo{Interestingly, the relation against mass computed using weak lensing has been shown to more closely follow self-similarity \citep{Mauro:2020}.}

Moving to panel (b), the $T_X-M_{500}$ relation in \simba\ is seen to follow a power law over almost the entire mass range. The predicted power law follows the self similar scaling relation, deviating slightly towards a shallower slope at lower masses ($M_{500}\la 10^{13.5}M_\odot$).  This again suggests that feedback mechanisms have the strongest impact in the poor group regime, increasing the temperature slightly over that expected from self-similarity.  The suppression in $L_X$ at these low masses thus does not owe to a suppression of the gas temperature, but instead to a lowering of the hot gas density.  The observations of $T_X-M_{500}$ suggest more deviation from self-similarity and hotter temperatures than predicted by \simba, though it is still mild \todo{, with recent observations from \citet{Umetsu:2020}showing a closer relation to self-similarity than previous observations, when computing $M_{500}$ through weak-lensing analysis}. Overall, however, \simba\ does a reasonable job of reproducing the $T_X-M_{500}$ relation.

In the lower left panel (c), we show the X-ray luminosity weighted iron abundance versus $M_{500}$ predicted in \simba. We will compare to observations when we examine the $\textrm{[Fe/H]}_{\odot} - T_X$ relation, but here we can already see that lower mass halos exhibit higher hot gas  metallicities on average, along with a larger scatter compared to the more massive halos. This suggests that in low mass halos, the hot gas is more associated with enriched feedback coming from within galaxies as opposed to more gravitational shock heating of more pristine infalling gas. By fitting a power law for halos with sSFR > -11 and sSFR < -11 we can see that halos with a lower specific star formation rate appear to exhibit a much steeper relation with lower mass halos having higher metallicities, while those halos with higher specific star formation rates have metallicities independent of $M_{500}$.  This is consistent with the idea that low-sSFR halos have been impacted by feedback energy that preferentially impacts smaller systems, and this feedback concurrently plays a role in enriching the hot gas.

Finally panel (d) shows the $S_{0.1}-M_{500}$ relation. Interestingly, $S_{0.1}$ shows almost no dependence on halo mass, and ranges between $10^2-10^3$~cm$^2$~K.  If anything, there appears to be a slight trend for higher entropies in lower mass systems, which again suggests a greater impact of non-gravitational processes on the hot gas at the poor group scale relative to more massive systems.  The observed groups from \citet{Pratt:2010} also show little trend, and in the overlapping mass range \simba\ produces fairly similar $S_{0.1}$ values.  The cause for the large spread is less clear, but given the very tight relation for $T_X-M_{500}$, it appears to be more related to variations in the core hot gas density. \todo{It is worth noting that several studies have found a hydrostatic mass estimate bias of $20-30\%$ \citep{Hurier,Linden:2014, Sereno:2017}, which would in turn bias any values for $M_{500}$ derived under the assumption of hydrostatic equilibrium, such as those from \citet{Eckmiller,Lovisari}, and \citet{sun9}. Accounting for such a bias by shifting \simba halo masses by $~20\%$  would see values exceed those of observations only in the most massive halos, where this bias is the weakest. Similarly a shift in the masses would see the \simba $T_X-M_{500}$ relation move closer to the observed trend.}

Overall, the scaling relations versus $M_{500}$ indicate a stronger impact from non-gravitational processes at $M_{500}\la 10^{13.5}M_{\odot}$, with a steepening of the $L_X-M_{500}$ relation, a slightly shallower $T_X-M_{500}$, and a mild increase in the hot gas metallicity and core entropy in the poor group regime.  \simba\ generally does reasonably well at reproducing available observations in amplitude and scaling with mass, albeit with somewhat too high $L_X$ values at high halo masses, and slightly too low $T_X$ at low halo masses.A detailed comparison using carefully constructed mock observations, analysed in a similar way, is required to assess how significant these discrepancies are; we leave this for future work.

\subsection{Scaling relations versus $T_X$}

\begin{figure*}
    \centering
    \includegraphics[width = \textwidth]{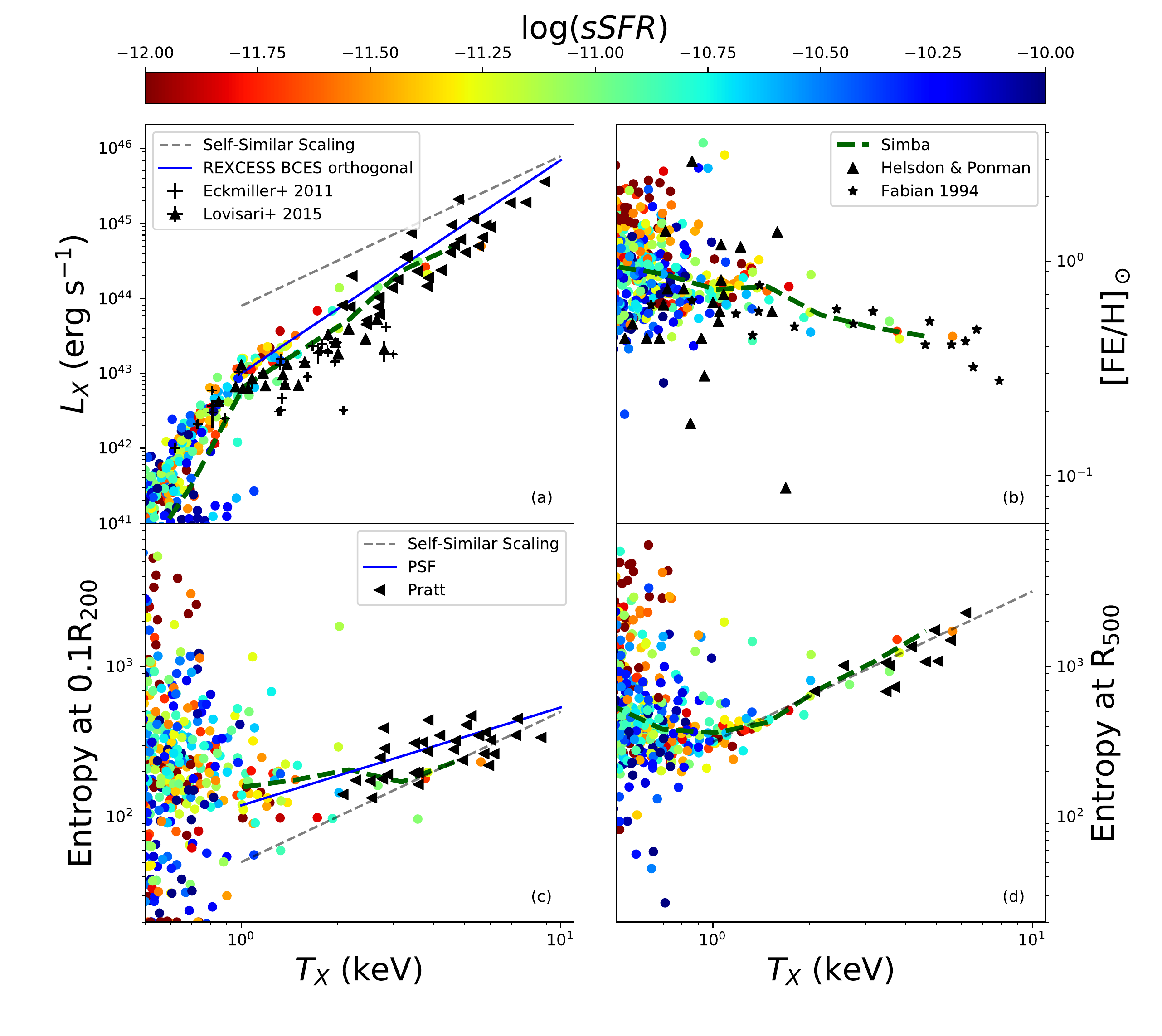}
    \caption{Scaling relations of X-ray properties versus $T_X$ in \simba at $z=0$, colour-coded by halo sSFR, with the dashed green line showing a running median. The figure shows the scalings of (a) $L_X$, (b) Metallicity, (c) $S_{0.1}$, and (d) $S_{500}$ versus $T_X$. Panel (a) shows observational data for the $L_X - T_X$ relation comes from REXCESS \citet{Pratt}, \citet{Eckmiller}, and \citet{Lovisari}, which \simba generally matching the relation by lying slightly above, likely owing to the slightly low temperatures seen in Figure~\ref{fig:Mass_Scaling}.  Panel (b) shows  observations from \citet{Helsdon} and \citet{Fabian94} scaled to match the solar abundances of \simba.  \simba shows a broadly similar trend of slightly increasing metallicities at lower $M_{500}$, but does not reproduce some of the data that shows quite low metallicities.  Panel (c) shows the best fit observed $S_X - T_X$ relation from \citet[PSF]{PSF}, along with data from \citet{Pratt:2010}.  \simba core entropies lie slightly below the observations at high-$T_X$, but below $T_X\la 2$~keV show an increase that deviates from self-similarity.  Panel (d) shows the entropy at $R_{500}$ and data from \citet{Pratt:2010}, which follows self-similarity very well down to $T_X\sim 1$~keV.
    Overall, \simba is broadly successful in reproducing available observations, with some small discrepancies.}
    \label{fig:Observational_scalings}
\end{figure*}

We now compare to relations versus the X-ray luminosity-weighted temperature $T_X$.  Due to the relative ease of observing $T_X$ as compared to halo mass, these relations are more commonly quoted by observers, so here we focus more on comparing to observations.

Figure \ref{fig:Observational_scalings} shows the scaling relations of various X-ray properties against $T_X$. The top left panel (a) shows $L_X - T_X$, with black points showing observational data from \citet{Pratt}. The top right panel (b) shows the X-ray luminosity weighted iron abundance [Fe/H]$_\odot - T_X$, with black markers representing observational data from \citet{Helsdon} and \citet{Fabian94}. The bottom left panel (c) shows $S_{0.1} - T_X$ with the observational line of best fit taken from \citet{PSF}. The bottom right panel (d) shows $S_{500} - T_X$ where $S_{500}$ is the entropy at $R_{500}$. The self similar relation for $L_X-T_X$ and both entropy relations are represented by the dashed black line. 

The first and best studied scaling relation is the $L_X -T_X$ relation \citep[e.g.][]{Mitchell,Markevitch,Maughan2012}.  Several studies have shown that this relation deviates from self similarity, suggesting that the gas heating is not due to gravitational processes alone. These studies have generally found that lower mass galaxy groups have increasingly larger deviation towards lower $L_X$ versus self similarity at $kT \la 2$~keV ( e.g. \citet{ponman1996rosat}, \citet{balogh1999differential}, \citet{Maughan2012}). It has been suggested that the deviation is as a result of the variation of gas content with mass. Since $L_X$ is proportional to the square of the gas density, this would drive a lowering of the observed luminosity for low mass systems, leading to a steeper relation.

In Figure \ref{fig:Observational_scalings} (a) we  see that \simba's $L_X-T_X$ relation follows a reasonably tight power law with a best-fit slope of 2.85 at $T_X\ga 1$~keV, steepening significantly below this.  There is also an increase in scatter at low $T_X$.  At all masses, this relation is steeper than the slope of 2 predicted for self-similarity, just as for the $L_X-M_{500}$ relation.  This indicates that \simba's implemented feedback mechanisms are working to alter the hot halo gas properties away from that shown by purely gravitational collapse, in a manner that is broadly concordant with observations.

The relations between entropy and temperature  provide a useful way to reflect on the $L_X - T_X$ relation. Processes that change the entropy of gas within the system will change how the gas behaves within it. For example, increasing the entropy of the gas would prevent gas from concentrating in the centre of the potential well, thereby reducing $L_X$.  AGN feedback, radiative cooling, star formation, and galactic winds are capable of heating the gas or removing low entropy gas thus effectively increasing the entropy in clusters.  This can result in the observed deviations from self similarity. These effects will act more strongly on low mass galaxies owing to their smaller potential wells, resulting in the observed trend of the most massive galaxies staying closer to self similar scalings. 

Investigating this within \simba we can see in Figure \ref{fig:Observational_scalings}(c) that the $S_{0.1} - T_X$ relation matches closely to the observed trend from \citet{PSF} across a  range of temperatures, showing a shallower relation than seen for self-similarity (dashed black line). In contrast, inspecting the entropy at $R_{500}$ ($S_{500}$), as shown in Figure \ref{fig:Observational_scalings}(d), we see a relation closer to what is seen in self-similar scaling. This would suggest that \simba's feedback is playing a larger role at 0.1$R_{200}$ closer to the central galaxy, while gravitational heating effects becoming dominant in the outskirts. Below 1~keV we see a large scatter in entropy at both 0.1$R_{200}$ and $R_{500}$. These halos have quite small amounts of hot gas to begin with, and thus can be more significantly impacted by energy injection, even out to $R_{500}$.  This suggests that while some small halos follow the trends seen at higher temperatures, feedback may be expelling low entropy gas in many cool halos resulting in increased entropy.  It is also noticeable that the halos with lower sSFR (redder points) tend to have higher $S_{500}$, and since jet feedback tends to occur in quenched galaxies, this corroborates the idea that jet feedback is responsible for the high entropies (i.e. low hot gas densities) in the outskirts of low-mass halos.

Figure \ref{fig:Observational_scalings} (b) shows the trend of $T_X$ with hot gas iron abundance.  There is a slight negative correlation between [Fe/H]$_{\odot} - T_X$ that generally matches observations across the range of observed temperatures.  At low temperatures, there is a larger scatter, but it appears that \simba tends to overproduce the hot gas metallicity.  This could be because the hot gas in these systems is more strongly impacted by jets from the central galaxy, which is carrying out metals.  Alternatively, these metals may have been deposited at an earlier epoch owing to star formation-driven winds, but now is being heated by jets.  Investigating this will require tracking individual particles to understand where and when metal injection occurs, which we leave for future work.

Overall, the temperature scaling relations demonstrate that \simba\ does a creditable job at reproducing various scaling relations, which is encouraging since no attempt was made to tune \simba\ to these observations.  There is a clear trend of non-gravitational feedback processes having a larger impact on hot gaseous halos at lower temperatures.  A potential discrepancy is the over-enrichment of hot gas in low-mass halos.  At $T_X\ga 1$~keV, \simba\ produces a steeper-than-self-similar relation versus $L_X$, a shallower-than-self-similar relation versus $S_{0.1}$, while showing self-similarity in $S_{500}$.  There is a blowup of the scatter in scaling relations at $T_X\la 1$~keV owing to the dominance of feedback energy in generating hot gas in these systems, something that could be tested with future observations with {\it Lynx} and {\it Athena}, or perhaps even {\it eROSITA}.  

\subsection{X-ray luminosity versus stellar mass}

Another interesting test of \simba\ is the relationship between the halo X-ray gas and the central galaxy.  This has been recently quantified in observations via stacking the X-ray halos around galaxies down to relatively low stellar masses \citep{Anderson:2015}. By using a stellar mass selection rather than X-ray selection, owing to the relatively tight relationship between halo mass and stellar mass \cite[e.g.][]{Behroozi:2013,Moster:2013}, one can probe the typical X-ray properties in such low mass systems. Given that many predicted X-ray properties in \simba\ seem to have larger scatter at lower masses presumably owing to the greater impact of feedback on hot gas, comparing to the \citet{Anderson:2015} data offers a novel approach to testing this regime.

\begin{figure}
    \centering
    \includegraphics[width = 0.49\textwidth]{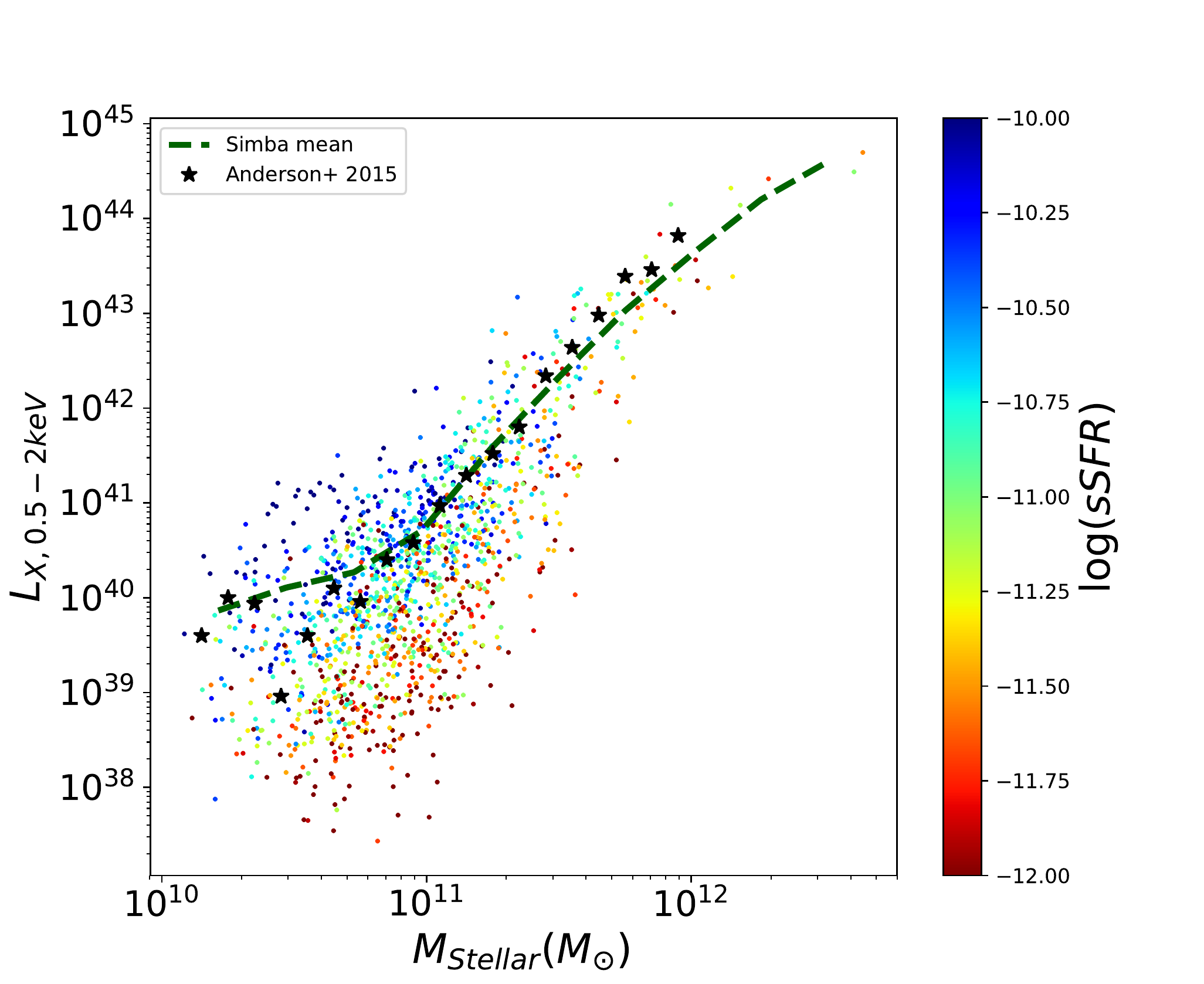}
    \caption{The scaling relation of central galaxy stellar mass $M_*$ versus \todo{total} x-ray luminosity calculated in the $0.5-2keV$ energy range, colour-coded by central galaxy sSFR and a running mean as the dashed green line. Stacked $L_X-M_*$ observations from \citet{Anderson:2015} are plotted as stars.  \simba shows quite good agreement with the observations down to the lowest $M_*$ values, including a flattening of the relation below $L^\star$, however \simba does sit slightly higher than observations in this region. \simba further shows a clear trend of star-forming galaxies having higher $L_X$ at a given $M_*$, but only at low masses since high-mass star-forming galaxies are rare. }
    \label{fig:LxMS} 
\end{figure}

Figure \ref{fig:LxMS} shows the stellar mass of central halo galaxy plotted against host halo X-ray luminosity.  Here we use the $0.5-2$~keV band to compute $L_X$, to match what was done in the stacked observational points taken from \citet{Anderson:2015} (shown as stars).  The dashed green line shows the running mean \twodo{of the linear $L_X$} from \simba, which is equivalent to what would be obtained for an analogous stacking exercise in \simba. We further colour-code the points by the sSFR of the central galaxy.

\simba\ follows the observed mean relation remarkably well over the full mass range probed from $M_*\approx 10^{10-12} M_\odot$. \twodo{\simba sits slightly above observations in the region below $L^\star$, and slightly below the most massive halos} but this is likely within the systematic uncertainties of this relatively crude comparison. This out-of-the box agreement over the full mass range is an impressive success of \simba.

Interestingly, there is no steepening of the $L_X-M_*$ relation at low stellar masses, as there is in the $L_X-M_{500}$ relation.  Instead, if anything there is a flattening of the relation at $L_X\approx 10^{40}$erg~s$^{-1}$ for low $M_*$, which is seen consistently in both the observations and in \simba.  \todo{ It is not immediately evident why there is a flattening; in this regime, X-ray binaries may begin to contribute significantly, but our analysis does not include these and yet \simba\ reproduces this flattening.}  This $M_*$ regime is where quenching must occur for the bulk of the galaxy population, so perhaps it is indicative of increased AGN feedback activity.  We will more fully explore the impact of AGN and other forms of feedback in heating halo gas in future work.

Finally, we note a strong trend versus sSFR: central star-forming galaxies tend to have larger X-ray luminosities than quenched ones at a given mass.  In \citet{Thomas:2019} we showed that star-forming galaxies in \simba\ often have substantial black hole accretion rates and thus black hole feedback, though typically not in jet mode.  It is possible that such feedback is still providing substantial heating of halo gas, particularly close to the galaxy, even if it does not strongly impact galaxy properties~\citep{Simba}.\todo{  A similar correlation between SFR and the $L_X-M$ relation was seen by \citet{Davies:2019} in EAGLE, although to a lesser extent, with the link believed to be due to black hole mass.}

Overall, the good agreement of \simba\ predictions with $L_X-M_*$ data suggests that \simba\ is properly capturing the interplay between feedback processes and gravitational heating processes around central galaxies at a wide range of stellar masses.

\section{AGN feedback variants}\label{sec:variants}

\begin{figure}
    \centering
    \includegraphics[width = 0.5\textwidth]{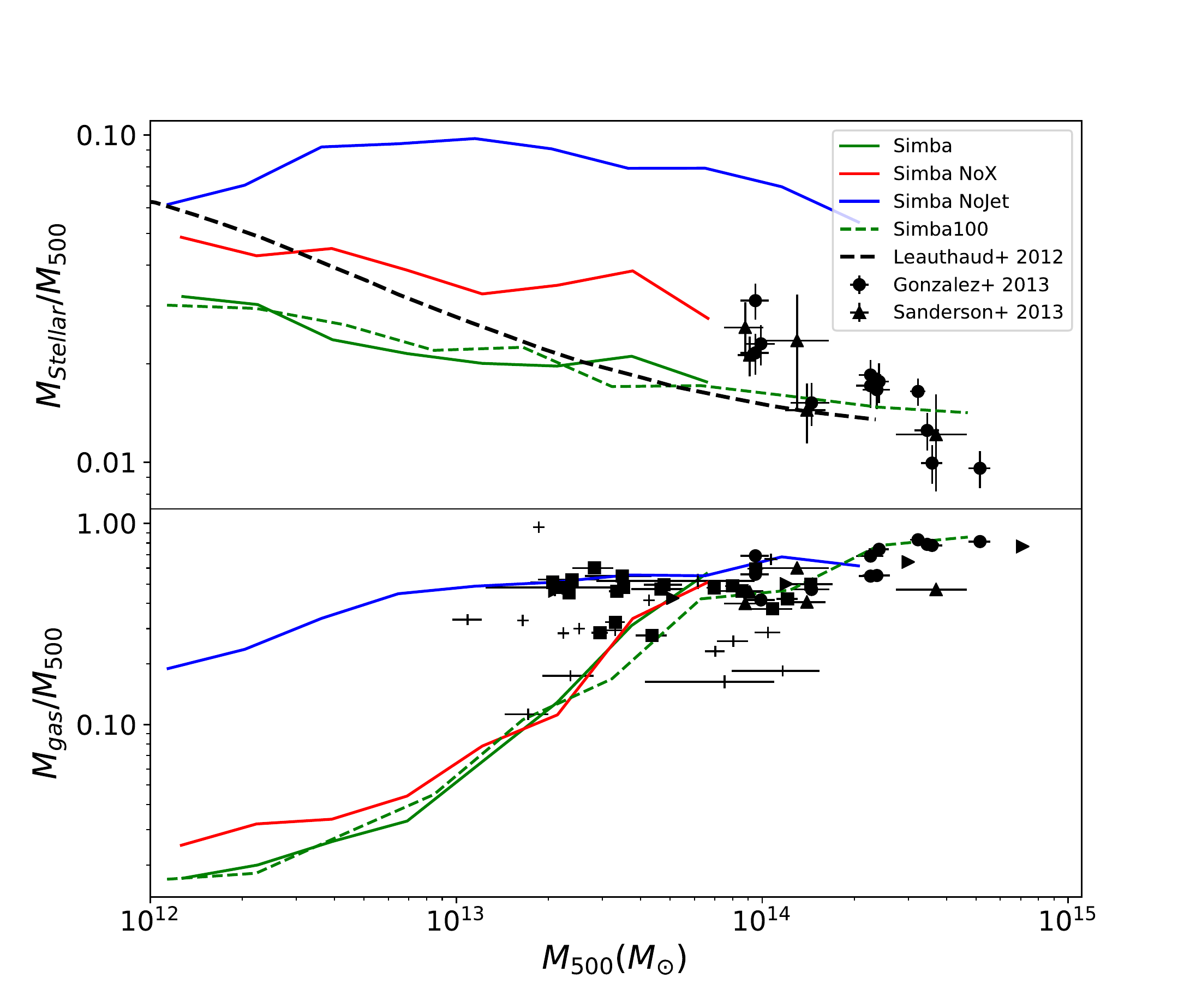}
    \caption{Running medians for the stellar mass fraction and hot gas mass fraction against $M_{500}$ for halos in the $50 h^{-1} Mpc$ boxes of \simba, \simba NoX, \simba NoJet, and \simba $100 \hmpc$. Jet feedback plays a distinct role in both the stellar, and hot gas mass fractions, bringing them in line with observations shown in Fig \ref{fig:Mass_Scaling}.}
    \label{fig:Fracs}
\end{figure}

\begin{figure}
    \centering
    \includegraphics[width = 0.5\textwidth]{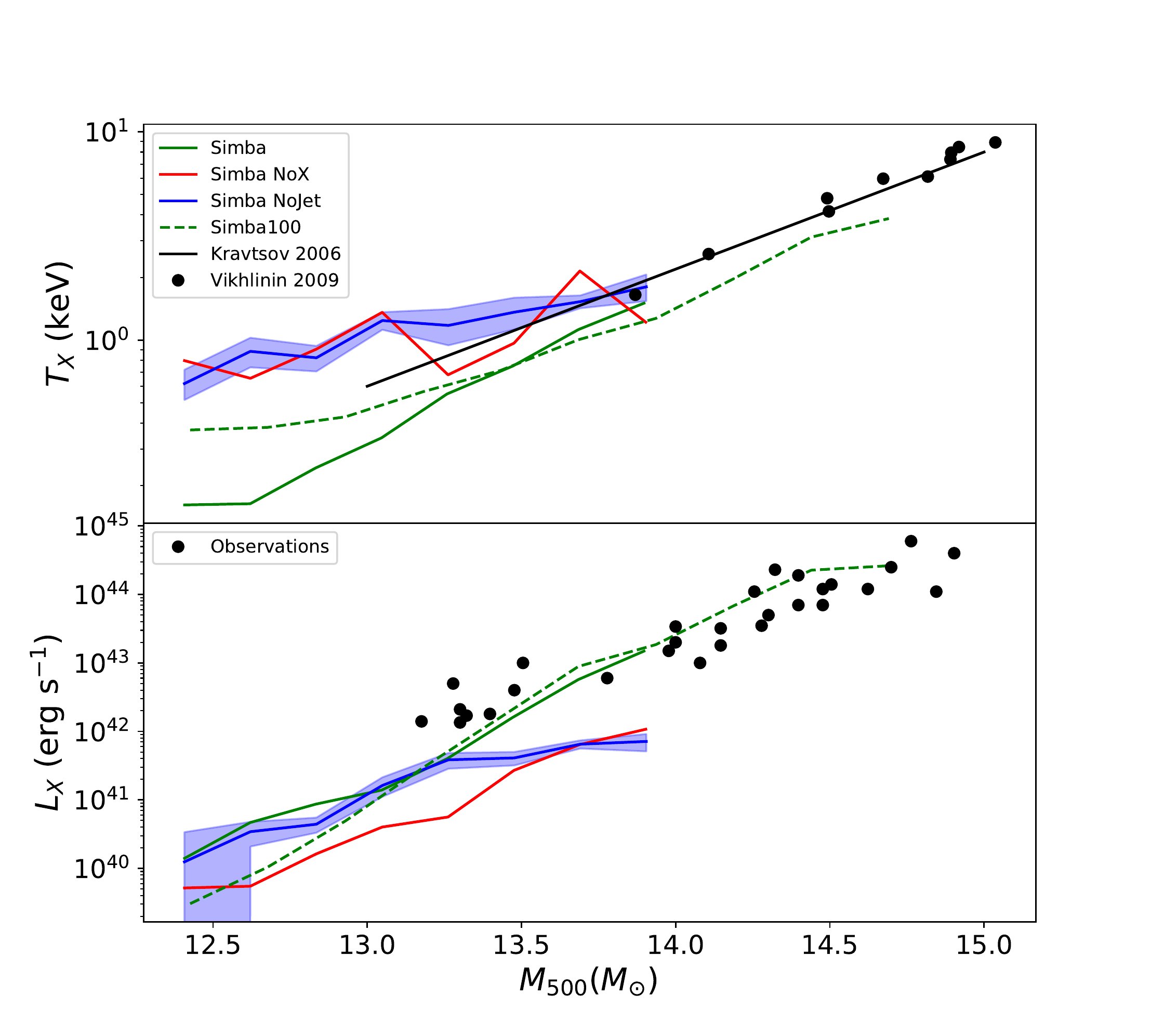}
    \caption{Running medians for the $T_X-M_{500}$ and $L_X-M_{500}$ scaling relations of halos in the $50 h^{-1} Mpc$ boxes of \simba, \simba NoX, \simba NoJet, and Mufasa, and the $100 h^{-1} Mpc$ box of \simba. A clear difference is seen in the $T_X-M_{500}$ relation at low masses, converging towards higher masses. \simba matches the $L_X-M_{500}$ more closely with observations than all other models suggesting the inclusion of X-ray feedback is working to increase luminosities, towards observed levels in the most massive halos.}
    \label{fig:All_Comp}
\end{figure}

An advantage of galaxy formation simulations is that it is possible to vary the included physics to directly quantify the impact of particular input physics modules.  To do this, we run versions of \simba\ with aspects of the AGN feedback turned off. We can thus compare full \simba with all feedback on, versus runs with the AGN jets and X-ray heating turned off (\simba-NoJet), and with only the X-ray heating turned off but AGN jets on (\simba-NoX).  Due to the computational time required, comparisons are made between various versions of \simba within $50\hmpc$ boxes, with $2\times 512^3$, thus providing the same resolution but one-eighth the volume.  This limits our statistics particularly at the massive end, but allows us to investigate the effect that each aspect of feedback has on the X-ray scaling relations and profiles in \simba. 

Figure~\ref{fig:Fracs} shows the median stellar-to-halo mass ratio (top) and the hot gas-to-halo mass ratio (bottom) as a function of $M_{500}$ in our three \simba\ variants, as well as \simba in the $100 \hmpc$ box (dashed green line).  For reference we reproduce the observations as shown in Figure~\ref{fig:StellarMass}, but our focus here is comparing between models.

The inclusion of AGN jet and X-ray feedback both have a significant impact on the stellar content in these massive halos.  The larger effect comes from jets (going from NoJet$\to$NoX,i.e. blue to red), but there is a significant effect when turning on the X-ray feedback as well (NoX$\to$\simba, i.e. red to green).  The jet feedback is primarily responsible for quenching galaxies~\citep{Simba}, but X-ray feedback provides important suppression of residual star formation in massive galaxies, by providing inside-out quenching~\citep{Appleby:2020}.  The impact is relatively invariant in mass, and including full AGN feedback physics tends to produce the best agreement with observations, at least in larger better-measured systems.

For the hot gas content, we also see  strong differences from including AGN feedback.  Here, X-ray feedback appears to have essentially no impact (green vs. red lines).  This is expected since X-ray feedback acts fairly locally to the black hole.  In contrast, AGN jets have a dramatic impact, strongly suppressing the hot gas content in low-mass halos.  This is better for producing agreement with observations suggesting low hot gas fractions in groups, although in the range where the models are strongly discriminated, the observations are currently inconclusive.

Finally, we note that these results are generally insensitive to simulation volume: The green dashed line from the $100\hmpc$ box follows the same trend as the $50\hmpc$ box.  This is a good sanity check that the smaller volume, while unable to produce the largest halos, provides an unbiased sample for the halos that it is able to generate.

To examine the effects of AGN feedback in terms of X-ray predictions, we look at both $T_X$ and $L_X$ against $M_{500}$.  Naively, one might expect that additional heating would increase the temperature, and would increase the entropy and thus lower the X-ray luminosity.

Figure \ref{fig:All_Comp} shows the comparison of $T_X$ and $L_X$ versus $M_{500}$ for the 3 AGN feedback variants of \simba.  The results for the full $100\hmpc$ \simba run is also shown (green dashed), which shows good agreement to the $50\hmpc$ \simba volume (green solid).  We show estimated cosmic variance for the NoJet case (shaded blue), by sub-sampling the median relation from 8 simulation sub-octants.  Observations of the $T_X - M_{500}$ relation from \citet{Kravtsov}, and \citet{Vikhlinin}  are shown. Observations of the $L_X - M_{500}$ relation come from \citet{Vikhlinin}, \citet{sun9}, and \citet{Pratt}. 

Counter-intuitively, the full feedback model as used in \simba\ significantly lowers the $T_X$ values across a large range of masses, with values beginning to match observations as well as the other models approaching $M_{500}\sim 10^{14}M_\odot$.   Turning on the jets (blue$\to$red) actually has very little impact on $T_X$, but turning on X-ray feedback (red$\to$green) significantly lowers the temperature.  Our implementation of X-ray feedback pushes dense gas outwards, which can add cooler gas to the core of the halo.  This also can add density to the core that will enhance cooling.  Note that with jets on, more baryons are evacuated, which lowers $M_{500}$; hence the shift in $T_X-M_{500}$ cannot be explained by changes in $M_{500}$ between the runs.

Meanwhile, for $L_X$, the runs are similar at low masses, while the jets increase $L_X$ in the more massive systems.  Comparing the NoX and Nojet cases, it appears that including jets lowers $L_X$, presumably by evacuating gas from the halos.  However, including X-ray feedback raises $L_X$ particularly in massive systems, since as argued above it increases the density of the core gas.  Note that since these boxes are all run from the same initial conditions, these variations, while small, cannot be explained by cosmic variance.

In general, these comparisons highlight the role of AGN jet feedback in establishing the properties of hot gaseous halos in groups.  They results in a strong reduction of the halo hot gas content particularly at low masses relative to a model without jets. The X-ray luminosities are not much impacted, but the X-ray temperatures are significantly lowered.  

\section{X-ray Profiles}

\subsection{Profiles as a function of mass}

\begin{table}
    \centering
    \caption{Median values at $R_{200}$ for $n_e$, L$_X$, T$_X$, S$_X$, binned by $M_{500}$.}
    \begin{tabular}{c c c c c }
     & $n_e$ &  L$_X$ & T$_X$ & S$_X$ \\
     $12.5 \leq M_{500} < 13$& $9.7 \times 10^{-6}$ & $2.1\times 10^{38}$ &  $0.29$ & $855.8$\\ 
     $13 \leq M_{500}< 13.5$&  $1.8 \times 10^{-5}$ & $4.3 \times 10^{39}$ &$0.31$ &$454.2$ \\ 
     $13.5 \leq M_{500}< 14$&  $3.3 \times 10^{-5}$ & $8.6 \times 10^{40}$ & $0.49$ &$411.6$\\
     $14 \leq M_{500}$&   $5.8 \times 10^{-5}$ & $1.1 \times 10^{42}$  & $0.87$   & $513.7$   \\
      
    \end{tabular}
    
    \label{tab:R200}
\end{table}

\begin{figure*}

    \centering
    \includegraphics[width = 0.85\textwidth]{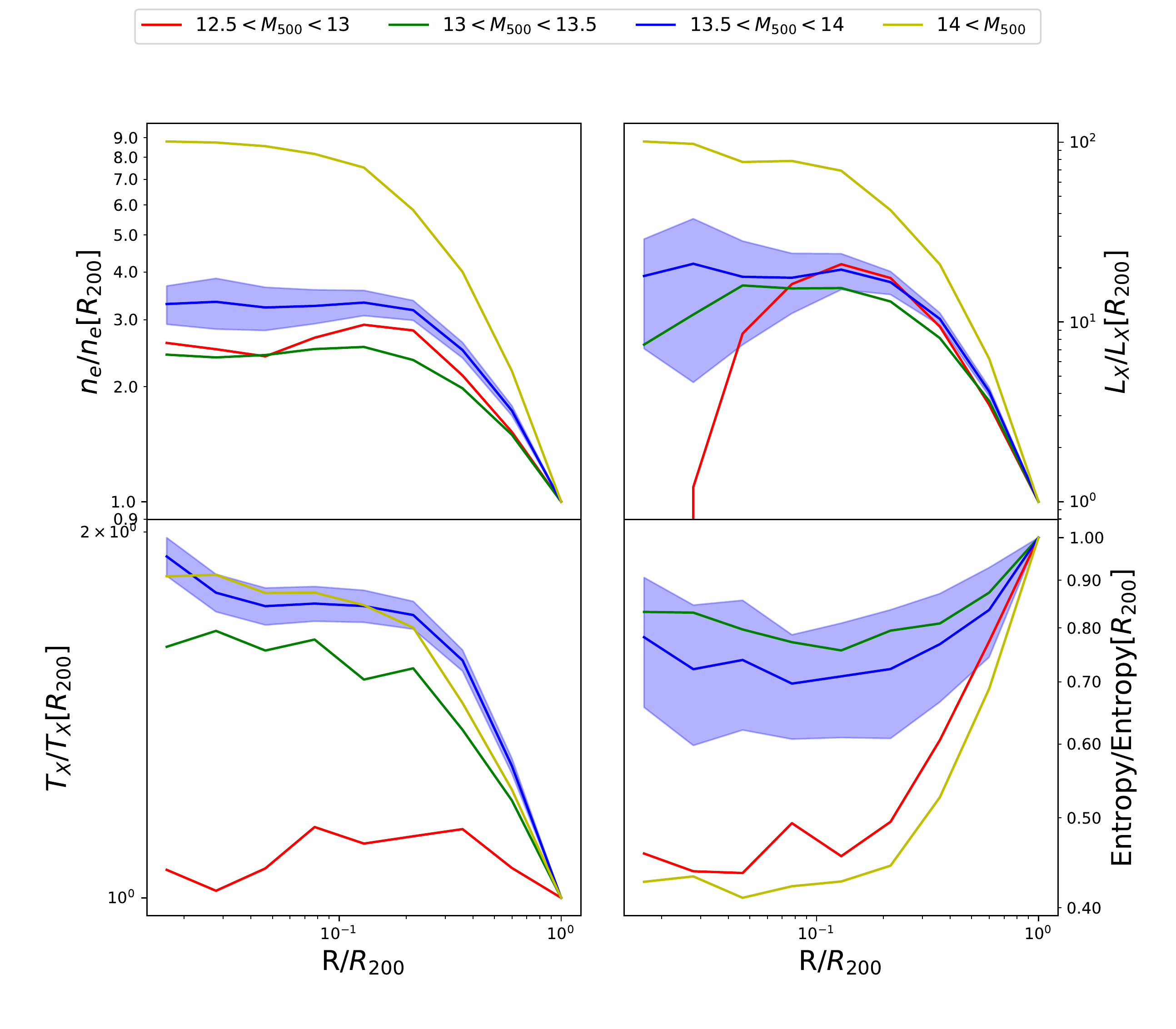}
    \caption{ Median profiles in bins of $12.5 \leq M_{500} < 13,13 \leq M_{500}< 13.5,13.5 \leq M_{500}< 14,14 \leq M_{500}$ for a) electron Density, b) $L_X$, c) $T_X$, and d) entropy, all scaled by the value at $R_{200}$ from top left to bottom right respectively.  The values at $R_{200}$ are listed in Table~\ref{tab:R200}.  Median values within radial bins are taken from the individual profiles of each halo, with the standard deviation plotted for the second most massive bin (which is representative) as the shaded blue region.  The electron density profile shows a core out to $\sim 0.2R_{200}$, and drops beyond that, more rapidly for the high-mass halos.  The luminosity profiles closely follow the shape of the electron density profiles, with $M_{500} \ga 10^{14}$ showing peaked core luminosities, while lower mass halos exhibit much flatter profiles. The $T_X$ profiles are fairly similar for all except the least massive groups, which show significant cooling in the core.  The entropy profiles of intermediate mass halos are considerably flatter than those of the most and least massive halos. 
    }
    \label{fig:Median_Profiles}
\end{figure*}

X-ray profiles provide a complementary view of how energy is deposited into hot gas owing to gravitational shock heating and non-gravitational processes.  To quantify profiles, we compute median values of stacked halo X-ray quantities from all halos in several mass bins above $10^{12.5}$ M$_{\odot}$, in log radial bins from $0.01R_{200}$ to $R_{200}$.  Electron density, $n_e$, is taken directly from the gas particles; $L_x$ is calculated as described in section \ref{Xray}; $T_X$ is given by the particles' temperatures weighted by $L_X$; and finally $S_X$ is calculated as $T_X / n_e^{2/3}$.  We also compute the $L_X$-weighted iron metallicity, scaled to a solar value of [Fe/H]$_\odot+12=7.50$~\citep{Asplund:2009}.

Figure \ref{fig:Median_Profiles} shows the median $L_x$, $T_x$, entropy, and electron density profiles for all halos with a $\log(M_{500}) > 12.5 M_{\odot}$, split into mass bins of $12.5 \leq \log(M_{500}) < 13,13 \leq \log(M_{500})< 13.5,13.5 \leq \log(M_{500})< 14,14 \leq \log(M_{500})$.  The profiles are normalised to their respective properties value at $R_{200}$ in order to better see the mass dependence of the profile shapes. The shaded blue region represents the standard deviation of the second most massive bin (where there are good statistics), divided by the total number of halos within said bin.  

The normalisation values at $R_{200}$ in these mass bins are presented in Table~1. Overall, this shows the expected trends that $n_e$, $L_X$, and $T_X$ all increase with increasing $M_{500}$.  However, $S_X$ shows a more interesting behavior, which we discuss after presenting the profiles.

In Figure \ref{fig:Median_Profiles}(a) we see the electron density profiles are similar across lower mass bins, but the most massive halos have a clearly higher electron density towards the central region. This is likely due to feedback having a more significant effect on lower mass halos, adding more energy with pushes more gas towards the outskirts of the halo, producing a flatter profile.

Panel (b) shows that the X-ray luminosity surface density largely follows electron density, with the halos of $M_{500} > 10^{14}M_\odot$ exhibiting a much more centrally peaked luminosity, with halos of $M_{500} < 10^{14}M_\odot$ showing a much flatter luminosity profile.  However, the lowest mass halos show a clear deficit of X-ray luminosity in the central region.  Since the electron density isn't dropping, this must owe primarily to the gas temperature.

This is seen in panel (c), which shows the temperature profiles.  Halos above $M_{500} > 10^{13}M_\odot$ show similar temperature profile shapes, dropping by a factor of two from the core to the outskirts, but the lowest mass halos show no temperature gradient. 

Finally in panel (d) we see the culmination of the two significant trends in $n_e$ and $T_X$. While the entropy in intermediate mass bins remains relatively flat with a slight increase towards $R_{200}$, the most and least massive bin see $\sim\times 2$ drop in entropy towards the core, caused by a steeper electron density profile, and a flatter temperature profile respectively.  Hence while the most massive halos are less affected, the intermediate mass range  shows a more significant impact from feedback, which means they may be good targets to examine the impact of AGN feedback. 

Returning to the normalisation values at $R_{200}$ from Table \ref{tab:R200}, we now discuss the entropy values that reveal a non-trivial trend. When compared to intermediate mass halos, the most and least massive halos have lower entropy at the core, rising much higher towards $R_{200}$ with the lowest mass halos showing significantly higher entropy towards the edges. While this doesn't follow the expected relation, by looking at Figure \ref{fig:Mass_Scaling} we can see that the large scatter at low masses is likely responsible, with entropy increasing at low temperatures (and as a result masses as seen in figure \ref{fig:Mass_Scaling}b) as we move outwards, shown in Figure \ref{fig:Observational_scalings}d.  For the most massive halos, $>10^{14} M_{\odot}$, the increase in entropy towards the outskirts further demonstrates the move from a flatter scaling relation at the centre towards self similarity at $R_{500}$ and $R_{200}$.

\begin{figure}
    \centering
    \includegraphics[width = 0.5\textwidth]{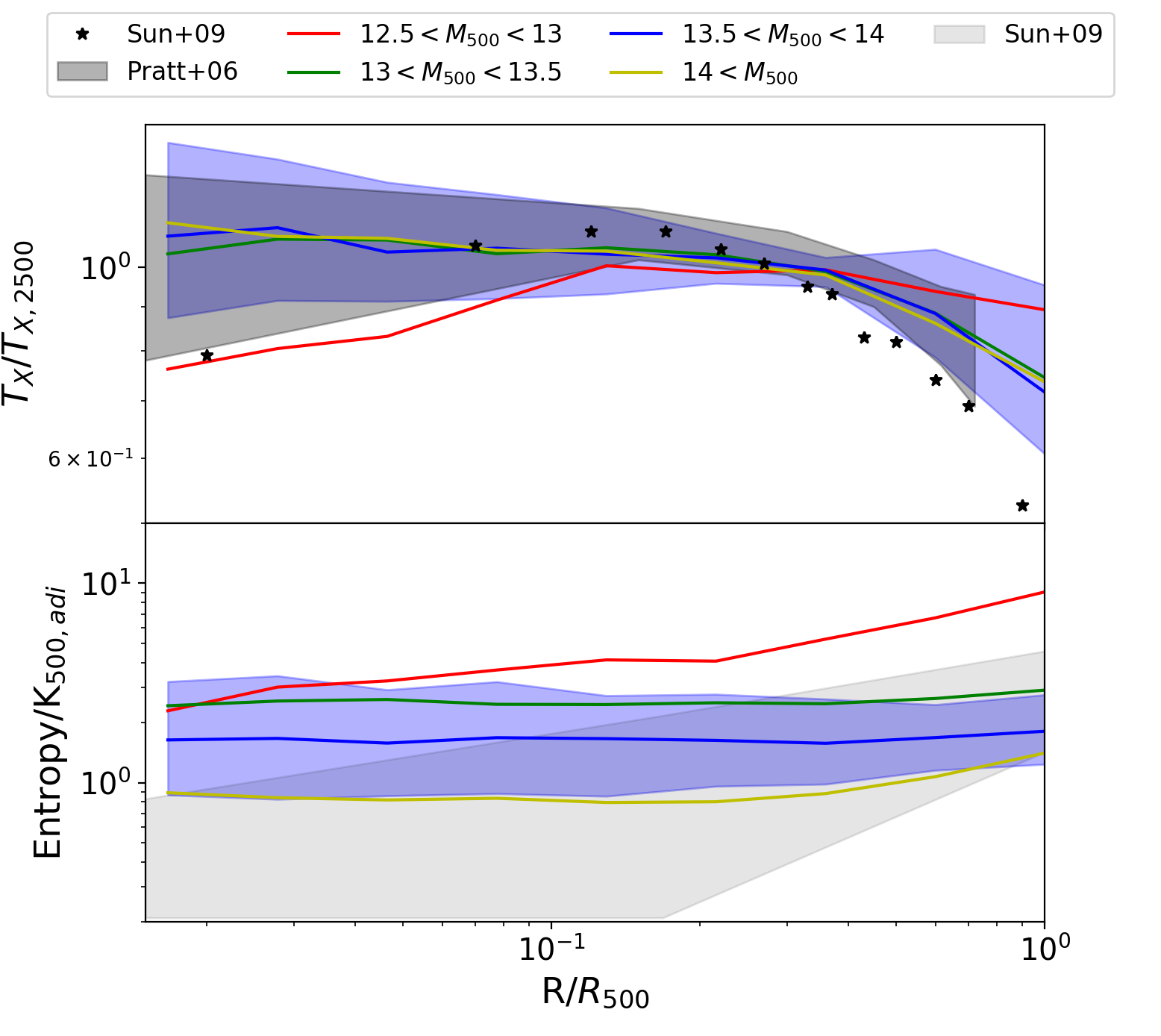}
    \caption{Median profiles of $T_X$ scaled by $T_{X,2500}$, and Entropy scaled by K$_{500,adi} $, binned by halo $M_{500}$ values. Observational data of the temperature profiles come from \citet{sun9} represented by the black stars,  and \citet{Pratt_2006} represented by the grey shaded region. A baseline entropy is provided by \citet{voit2005expectations}, while the upper envelope on entropy profiles comes from \citet{sun9}.}
    \label{fig:observationProfile}
\end{figure}

\todo{In Figure \ref{fig:observationProfile} we compare both temperature, and entropy profiles to observations. In order to do this we must first scale the profiles to be comparable to observational data. Temperature profiles from \citet{sun9} are scaled by $T_X$ within $R_{2500}$, allowing us to easily scale to this value. Profiles from \citet{Pratt_2006} are scaled by $T_{vir}$ and scaled to $R_{200}$.  The flat cores seen within \simba mean that scaling $T_{vir}$ or $T_{X,2500}$ makes little difference in the most massive halos, however the least massive halos do experience a shift above observed profiles when scaling by $T_{vir}$.  }

\todo{The halos of comparable mass to the data (i.e. the intermediate mass bins of $10^{13-14}M_\odot$) closely follow observations from \citet{Pratt_2006}, however show a  flatter core than \citet{sun9}, and perhaps also a flatter outer profile out to $R_{500}$. These higher temperatures in the core regions suggest that the feedback may be too efficient in heating the inner regions. }

\todo{This inner heating is also reflected in the entropy profiles.  To compare to observations, we scale the entropy profiles by the adiabatic entropy scale $K_{500,adi}$, as described in \citet{sun9}. While there is a large scatter in the entropy profiles in the group regime, we denote by the shaded region the area between by baseline entropy profiles from \citet{voit2005expectations} at the lower end, and an estimate of the upper envelope on profiles shown in \citet{sun9}.  Profiles from \simba show a flatter entropy core, steepening towards $R_{500}$.  In fact, the profiles are in better agreement in our lowest mass bin, though the amplitude, relative to  $K_{500,adi}$, is elevated.}

\todo{These results suggest that feedback in \simba\ is over heating the inner regions of halos, producing too large an entropy core.  We note that the outer entropy profile at low and high masses do have a significant upturn, as can be seen in Figure~\ref{fig:Median_Profiles} which has a much smaller $y$-axis range, but at intermediate masses they are too flat.  This is in contrast to e.g. FABLE~\citep{Henden}, which shows entropy profiles in better agreement with observations, falling into the inner regions, though some of their profiles seem to show too steep a drop in the centre.  This demonstrates the power of using the inner profiles of hot gas in groups to constrain the details of AGN feedback injection.}

\begin{figure}
    \centering
    \includegraphics[width = 0.5\textwidth]{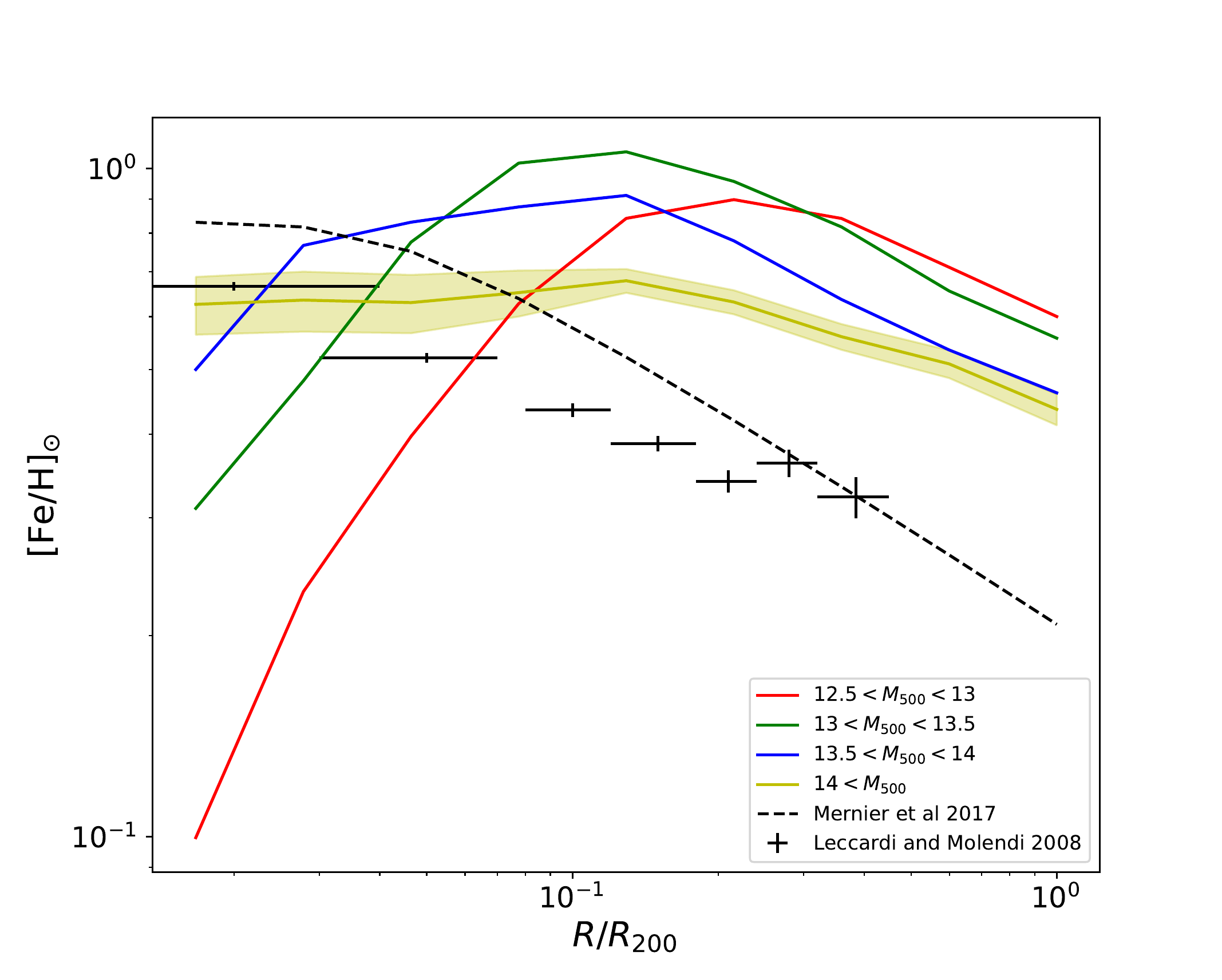}
    \caption{Median profiles of halo solar metallicity profiles in \simba, binned by the halo $M_{500}$ values. Observational data of the metallicity within $0.4 R/R_{200}$ are taken from \citet{Leccardi} and \citet{Mernier}. The X-ray weighted iron metallicity profiles are over predicted in all mass bins, agreeing with the $[FE/H]_{\odot} - M_{500}$ relation seen in figure \ref{fig:Mass_Scaling}.  } 
    \label{fig:Metallicity}
\end{figure}

Figure \ref{fig:Metallicity} shows the X-ray weighted iron metallicity profiles binned by $M_{500}$ as before. A fit to observations from \citet{Leccardi} and stacked X-ray data from \citet{Mernier} are plotted in black.  The yellow shading shows standard deviation around the largest mass bin values.

Some interesting trends are notable.  First, there is clear trend in the profile shapes with increasing $M_{500}$.  
The highest mass halos show a shallow declining trend with radius throughout.  These are probably the most comparable to the observed halo samples, though probably still somewhat lower masses.  Overall, however, it is clear that while the core metallicity agrees well with observations and thus leads to reasonable looking global X-ray luminosity weighted metallicities (as seen in Figure~\ref{fig:Observational_scalings}), the iron abundance in the outskirts is clearly well above observations.  This suggests overly widespread metal enrichment within the hot halo gas.  It is possible that differences in the typical halo mass between the observations and \simba\ could be partly responsible, but this would require next-generation X-ray facilities to examine.  For now, we view this as a failing of \simba\ that should be investigated, perhaps via particle tracking to identify where the iron in the outskirts originates.

At lower halo masses, the metallicity profiles are significantly higher in the outskirts, in part reflecting the larger spread in metallicities seen in Figure \ref{fig:Mass_Scaling}. This suggests that at $R\ga 0.1R_{200}$, the hot gas has been considerably more enriched than in higher mass systems.  This is consistent with the idea that we have seen before, that the hot gas in low-mass systems is more strongly dominated by feedback coming from the galaxy, which carries out enriched galactic gas.  Interestingly, at small radii, the hot gas metallicity in low-mass groups drops quickly. It is not immediately evident why this is, though it may be impacted by the fact that jets are decoupled from hydrodynamics (and excluded from our analysis while decoupled) out to as much as $\sim 0.1R_{200}$ in the smallest halos.  Particle tracking analyses can help probe the nature of metal distribution; we leave this for future work.

\todo{In summary, while \simba\ generally does a good job of reproducing overall group and cluster properties, comparisons to observed profiles identify some non-trivial discrepancies between \simba\ predictions and observations.  In particular, while the temperature profiles are in general agreement, the entropy profiles are too shallow, suggesting that \simba\ has displaced too much gas from the central regions, particularly in intermediate-mass groups.  The metallicity profiles, in contrast, show an over-abundance of iron in the outskirts, which may reflect too much enriched gas being moved from the core to the outer parts.  These suggest that \simba's feedback model may need adjustment to better match these constraints.  Improved observations from {\it eRosita} will hopefully provide larger and more robust samples for comparison.}

\subsection{The impact of AGN feedback on profiles}

To better understand the physics that shapes the hot gas profiles, we now examine these same X-ray profiles in the AGN feedback variant runs discussed in \S\ref{sec:variants}.

\begin{figure}
    \centering
    \includegraphics[width=0.5\textwidth]{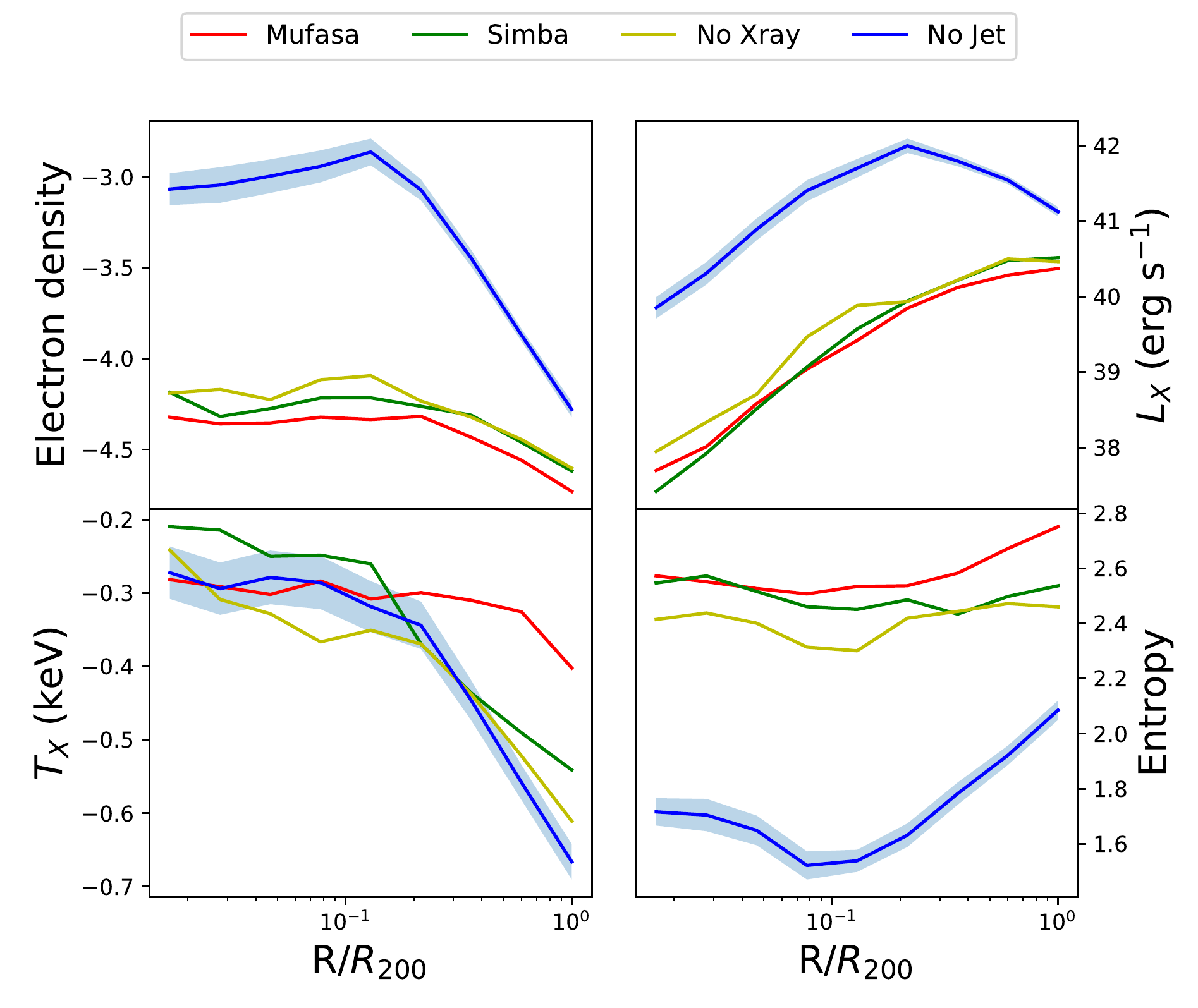}
    \caption{Electron Density, $L_X$, $T_X$. and Entropy profiles for halos with mass $ 10^{13} < M_{500} < 10^{14} $ within Mufasa, \simba, \simba NoX, and \simba NoJet. From these profiles we see that the inclusion of jet mode feedback has a significant impact on both the shape and value of the profiles.}
    \label{fig:Profile_Comparison}
\end{figure}

Figure~\ref{fig:Profile_Comparison} shows X-ray profiles in our three $50\hmpc$ AGN feedback variant runs discussed in \S\ref{sec:variants}.  In particular, we show median profiles of $n_e$, $L_X$, $T_X$, and $S_X$ as a function of $R/R_{200}$ for all halos with masses of $10^{13} \textrm{M}_{\odot} < \textrm{M}_{500} < 10^{14} \textrm{M}_{\odot}$, which is the mass range where AGN feedback appears to have the largest impact on X-ray properties.   Here we further compare with the older \mufasa\ simulation, which did not include black holes or associated feedback but rather used a halo gas heating scheme to quench galaxies.  \mufasa\ was run with the identical $50\hmpc$ initial conditions of the \simba\ variants.  Hence we are also able to see how \simba's new model of feedback impacts hot gas scaling relations.

Here, by far the strongest impact is seen from including jet AGN feedback.  The NoJet run has much higher electron densities and $L_X$ values, and greatly reduced entropies, compared to the runs with jets (either NoX or the full \simba\ case).  Hence, jets are crucial for raising the entropy of hot halo gas, which is directly related to its ability to quench star formation. It is interesting that despite the fact that jets are implemented in a purely bipolar fashion, the entropy increase is felt throughout the halo.

Interestingly, there is very little impact on the temperature profiles, though full \simba\ does have mildly higher temperatures in the inner regions apparently owing to X-ray feedback.  This shows that X-ray feedback primarily has an impact within $\la 0.1R_{200}$, which is as expected since it acts only on gas reasonably close to the black hole.  The entropy profiles are fairly flat in all cases, but strongly increased by the jet feedback. \todo{Thus it appears to be the jet feedback that causes the overly flat entropy profiles in \simba.}

The \mufasa\ run, despite its very different feedback mechanism, produces $n_e$, $L_X$, and $S_X$ profiles that are very similar to \simba's. \mufasa's preventive feedback model was an ad hoc shutting off (or offsetting) of cooling in diffuse halo gas above a (mildly) evolving mass threshold, tuned to quench galaxies.  \simba's jets, despite being implemented kinetically and in a bipolar manner, appears to have the same general effect on the physical properties of halo gas as a function of radius.  This is likely why it ends up having similar massive galaxy properties such as producing a observationally-concordant population of quenched galaxies, despite implementing quite different feedback schemes.

\section{Summary}
In this paper we have examined the X-ray scaling relations, and X-ray property profiles of halos within the \simba\ cosmological galaxy formation simulations.
Using hot X-ray emitting gas within groups and clusters we are able to discriminate between various models and place constraints on the physical processes governing AGN feedback and galaxy formation in dense environments. 
We have shown that:
\begin{itemize}
    \item \simba\ is able to produce good agreement with observed stellar baryon and hot gas mass fractions as a function of halo mass.  The hot gas fraction drops increasingly rapidly towards low halo masses, showing the importance of gas evacuation by feedback. \todo{However \simba sits on the low end of observations at intermediate masses indicating it may be under producing hot gas below $10^{13}\textrm{M}_{\odot}$ }
    \item \todo{Although} observed X-ray scaling relations against $M_{500}$ are broadly successfully reproduced, we do see $L_X-M_{500}$ falling beneath observations in the least massive halos. \todo{\simba halos are consistently cooler than observations across all masses by $<0.2$dex, but it is important to note that correcting for a potential underestimate of $M_{500}$ due to the assumption of hydrostatic equilibrium could bring \simba and observations in line.} Changes in these relations below $M_{500} \la 10^{13.5} M_{\odot}$ such as an increased slope for $L_X-M_{500}$, a $T_X$ deviating above self-similarity, and an increased scatter in [Fe/H] and entropy, $S_X$, indicate a stronger impact from non-gravitational processes in poor groups versus larger systems.
    \item \simba's predicted X-ray scaling relations versus X-ray temperature, $T_X$, broadly reproduce observations despite no attempt to tune the model to these data. These include $L_X-T_X$, [Fe/H]$-T_X$, $S_{0.1}-T_X$, and $S_{500}-T_X$.  
    At $T_X \leq 1 keV$, \simba\ predicts more rapidly dropping $L_X$, an increasing metallicity, and a large entropy scatter; these predictions can be tested with statistical samples from e.g. {\it eROSITA}.
    \item X-ray profiles of \simba\ halos vary significantly in shape at high and low masses.  In $M_{500}>10^{14}M_\odot$ halos, the electron density profile increases sharply down to $0.1R_{200}$ before flattening into a core, whereas lower mass halos have flat $n_e$ profiles.  Halos with $M_{500}>10^{13}M_\odot$ have dropping temperature profiles as expected for hydrostatic equilibrium, but the least massive systems show a flat $T_X$ profile, likely owing to efficient cooling in the core.  The entropy profiles are thus rising rapidly beyond $0.1R_{200}$ in high and low mass systems, but have a plateau at a higher value in intermediate mass halos.
    \item \simba's temperature profiles are in reasonable agreement with observations, but the entropy profiles are too flat in the inner regions, showing an overly-large core for intermediate mass halos.  This suggests that too much gas has been removed from the central region by \simba's feedback.
    \item \simba's X-ray weighted iron metallicity profile is too flat compared to observations, showing agreement in the core but significantly higher metallicities than observed in the outskirts.
    \item Using variants of \simba\ with different AGN feedback modules turned on and off, we show that \simba's jet AGN feedback is most responsible for altering the halo properties, primarily by evacuating hot gas and thus lowering the electron density.
    \item The \mufasa\ simulation, despite using a quite different approach to modeling AGN feedback, yields profiles similar to \simba.  This shows that the first-order impact of \simba's AGN jets is to keep halo gas hot via energy injection, as is done explicitly in \mufasa.  However, there are some differences such as the X-ray temperature at radii $\ga 0.3R_{200}$ being higher in \mufasa\ than in \simba.
    
\end{itemize}

Overall, \simba\ is successfully demonstrating the balance between feedback and gravitational heating in halo gas around massive galaxies across a range of halo masses. This is an independent test of \simba's feedback models, as compared to the stellar or black hole properties that have been examined in previous works.  The fact that there was no tuning explicitly done to match X-ray halo gas properties is a nice success of \simba's feedback model, and in particular its AGN jet feedback implementation.  Since the primary impact of AGN jets is to evacuate gas increasingly particularly in lower-mass halos, improving constraints on the hot gas contents of group-sized halos promises to be an important way to constrain models of AGN feedback.

The X-ray profiles highlight additional constraints on how feedback alters group and cluster X-ray properties.  In particular, \simba's jet feedback causes a strong drop in the electron density and thus the X-ray luminosity at all radii, moreso in lower mass systems.  The consequence is that the entropy of the hot gas is raised substantially, which then prevents gas cooling and resulting star formation which would otherwise happen more vigorously particularly in lower mass halos.  The solution to the cooling flow problem in \simba\ is thus provided by gas evacuation owing to AGN jet feedback.

\todo{However, the profiles also highlight significant discrepancies between \simba\ and observations.  The shallow entropy profiles in $M_{500}\la 10^{14}M_\odot$ systems suggests that gas is being redistributed too strongly from the core regions into the outskirts.  This may require significant modifications to the jet feedback scheme.  Relatedly,} 
the fact that the metallicity profile is too flat compared to observations raises a more significant issue.  In these systems, iron is generated by Type~Ia and Type~II supernovae, so the discrepancy may reflect something to do either with the way iron is generated  or how it is distributed from its various sources.  A complementary probe would be to examine $\alpha$ elements such as oxygen or magnesium, which arise purely in Type~II supernovae.  Good spectral data for low-mass halos is necessary to measure these emission lines, which would be a valuable constraint on X-ray gas enrichment processes.  At face value, the excess in iron in the outskirts suggests overly widespread dispersal from feedback, though it is not typically AGN feedback that causes significant metal dispersal since it does not carry much mass as compared to star formation feedback.

There is much future work to pursue in examining the properties and growth of hot X-ray gas in groups.  We plan to examine how X-ray halos develop, what additional constraints may be obtained on AGN feedback physics from evolutionary trends, and how next-generation facilities will shed new light on the physics of intragroup and intracluster gas. \simba\ provides a plausible platform to interpret such observations, which in turn will provide valuable constraints on uncertain input physics processes such as AGN feedback.

\section*{Acknowledgements}

The authors thank Weiguang Cui, Katarina Kraljic, and Sarah Appleby for helpful discussions, and Nick Henden for providing the data from the FABLE simulations.
RD acknowledges support from the Wolfson Research Merit Award program of the U.K. Royal Society.
This work used the DiRAC@Durham facility managed by the Institute for Computational Cosmology on behalf of the STFC DiRAC HPC Facility. The equipment was funded by BEIS capital funding via STFC capital grants ST/P002293/1, ST/R002371/1 and ST/S002502/1, Durham University and STFC operations grant ST/R000832/1. DiRAC is part of the National e-Infrastructure.

\section*{Data Availability}
Data will be made available on request. 

\bibliographystyle{mnras}
\bibliography{references.bib} 






\bsp	
\label{lastpage}
\end{document}